\documentclass[12pt]{article}
\pdfoutput=1

\usepackage{
color,
amssymb,amsmath,bm,bbm
}

\usepackage[frozencache,cachedir=.]{minted}

\usepackage{afterpage}
\usepackage{longtable}
\usepackage{booktabs}
\usepackage[dvipsnames]{xcolor}
\usepackage[linktoc=page,bookmarks=false,colorlinks=false,linkbordercolor=RoyalBlue,citebordercolor=ForestGreen,urlbordercolor=CornflowerBlue]
{hyperref}
\usepackage{latexsym,mathrsfs,dsfont}
\usepackage[compress]{cite}
\usepackage{graphicx}
\usepackage{url}
\usepackage{paralist}
\usepackage{bbold}
\usepackage{xspace}

\usepackage{amsmath}
\usepackage{slashed}
\usepackage{xparse}
\usepackage{etoolbox}

\newcommand{\skipp}[1]{}

\newcommand{\talk}[3]{
    \section{#1}
    \vspace{-3mm}
    \begin{flushright}\textit{Speaker: #2}\\#3\end{flushright}
}

\newcommand{\mrdim}{\mathrm{dim}}

\newcommand{\mrO}{{\mathrm{O}}}
\newcommand{\mra}{{\mathrm{a}}}
\newcommand{\mrA}{{\mathrm{A}}}
\newcommand{\mrS}{{\mathrm{S}}}
\newcommand{\mrE}{{\mathrm{E}}}
\newcommand{\SMEFT}{\rm{\scriptscriptstyle{SMEFT}}}
\newcommand{\eff}{{\mbox{\scriptsize{eff}}}}
\newcommand{\bqas}{\begin{eqnarray*}}
\newcommand{\eqas}{\end{eqnarray*}}
\newcommand{\nl}{\nonumber\\}

\def\be{\begin{equation}}
\def\ee{\end{equation}}
\def\bea{\begin{eqnarray}}
\def\eea{\end{eqnarray}}
\def\nnb{\nonumber}
\newcommand{\Lu}{\mathcal{L}}
\newcommand{\f}{\frac}
\newcommand{\scs}{\scriptscriptstyle}

\newcommand{\mm}{\texttt{MatchMaker}\xspace}
\newcommand{\mathematica}{\texttt{Mathematica}\xspace}
\newcommand{\form}{\texttt{FORM}\xspace}
\newcommand{\python}{\texttt{Python}\xspace}

\newcommand{\qgraf}{\texttt{QGRAF}\xspace}
\newcommand{\feynrules}{\texttt{FeynRules}\xspace}

\newcommand{\EOS}{\texttt{EOS}\xspace}

\newcommand{\DEFT}{\texttt{DEFT}\xspace}
\newcommand{\op}{\mathcal{O}}
\newcommand{\opB}{\mathcal{B}}

\newcommand{\SMEFTsim}{\texttt{SMEFTsim}}
\newcommand{\Feynrules}{\texttt{FeynRules}}
\newcommand{\MG}{\texttt{MadGraph5\_aMC@NLO}}
\newcommand{\wcxf}{\texttt{WCxf}}
\newcommand{\dimsixtop}{\texttt{dim6top}}
\newcommand{\SMEFTatNLO}{\texttt{SMEFT@NLO}}

\def\sfr{{\tt SmeftFR}~}
\def\frules{{\tt FeynRules}~}

\newcommand{\chpt}{$\chi$PT}
\newcommand{\nn}{\nonumber\\ }
\renewcommand{\O}{\mathcal{O}}
\renewcommand{\L}{\mathcal{L}}

\definecolor{darkred}{rgb}{0.6,0,0}

\newcommand{\pkg}[1]{{\tt #1}\xspace}
\newcommand{\dsix}{\pkg{DsixTools}}
\newcommand{\dsixo}{\pkg{DsixTools 1.0}}
\newcommand{\dsixv}{\pkg{DsixTools 2.0}}

\newcommand{\mathe}{\pkg{Mathematica}}
\newcommand{\json}{\pkg{JSON}}
\newcommand{\yaml}{\pkg{YAML}}

\begin{document}

\begin{flushleft}
{\Large\bf
\boldmath{
{\Huge Computing Tools for the SMEFT}
}}
\\[5mm]
{\small {\bf SMEFT-Tools 2019}, 12-14th June 2019, IPPP Durham}
\\[5mm]

{\it Editors:}\\[2mm]

{Jason Aebischer${}^a$, Matteo Fael${}^b$, Alexander Lenz${}^{c}$, Michael Spannowsky${}^{c}$, Javier Virto${}^{d}$} \\[3mm]

{\it Contributors:}\\[2mm]

{Ilaria Brivio${}^{e}$, Juan Carlos Criado${}^f$, Athanasios Dedes${}^g$, Jacky Kumar${}^{h}$, Miko{\l}aj Misiak${}^i$,\\
Giampiero Passarino${}^j$, Giovanni Marco Pruna${}^k$, Sophie Renner${}^l$, Jos{\'e} Santiago${}^f$,\\
Darren Scott${}^{m,n}$, Emma Slade${}^o$, Peter Stangl${}^p$, Peter Stoffer${}^a$, David M. Straub${}^q$,\\
Dave Sutherland${}^r$, 
Danny van Dyk${}^s$, Avelino Vicente${}^t$}\\[5mm]

{\scriptsize
${}^{a}$
Department of Physics, University of California at San Diego, La Jolla, CA 92093, USA\\
${}^{b}$
Theoretische Physik I, Universit\"at Siegen, 57068 Siegen, Germany\\
${}^{c}$
IPPP, Department of Physics, University of Durham, DH1 3LE, United Kingdom\\
${}^{d}$
Departament de F\'isica Qu\`antica i Astrof\'isica and ICCUB, Universitat de Barcelona, 08028 Barcelona, Catalonia\\
${}^{e}$
Institut  f\"ur  Theoretische  Physik, Universit\"at  Heidelberg,
Philosophenweg  16,
DE-69120  Heidelberg, Germany \\
${}^f$
CAFPE and Departamento de F\'isica Te\'orica y del Cosmos, Universidad de Granada,
E-18071, Granada, Spain\\
${}^g$
Department of Physics, University of Ioannina, GR 45110, Ioannina, Greece\\
${}^{h}$
Physique des Particules, Universit\'e de Montr\'eal,
C.P. 6128, succ.  centre-ville, Montr\'eal, QC, Canada H3C 3J7\\
${}^i$
Institute of Theoretical Physics, Faculty of Physics,University of Warsaw,
Pasteura 5, PL 02-093, Warsaw, Poland\\
${}^j$
Dipartimento di Fisica Teorica, Universit\`a di Torino, and 
INFN Sezione di Torino, Italy\\
${}^k$
Laboratori  Nazionali  di  Frascati,  via  E.  Fermi  40,  I–00044  Frascati,  Italy\\
${}^l$
SISSA International School for Advanced Studies, Via Bonomea 265, 34136, Trieste, Italy\\
${}^m$
Institute for Theoretical Physics, University of Amsterdam, Science Park 904,
1098 XH Amsterdam, The Netherlands \\
${}^n$
Nikhef, Theory Group, Science Park 105, 1098 XG, Amsterdam, The Netherlands\\
${}^o$
Rudolf Peierls Centre for Theoretical Physics, University of Oxford,
Oxford OX1 3PU, United Kingdom\\
${}^p$
Laboratoire d’Annecy-le-Vieux de Physique Th\'eorique, UMR5108, CNRS,
F-74941, Annecy-le-Vieux Cedex, France\\
${}^q$
Excellence Cluster Universe, Boltzmannstr. 2, 85748 Garching, Germany\\
${}^r$
Department of Physics, University of California, Santa Barbara, CA 93106, USA\\
${}^s$
Technische Universit\"at M\"unchen, James-Franck-Strasse 1, D-85748 Garching, Germany \\
${}^t$
Instituto de F\'isica Corpuscular (CSIC-Universitat de Val\`encia), Apdo. 22085, E-46071 Valencia, Spain\\
}

\end{flushleft}

\vspace{6mm}

\noindent
{\bf Abstract}\\
The increasing interest in the phenomenology of the Standard Model Effective Field Theory (SMEFT), has led to the development of a wide spectrum of public codes which implement automatically different aspects of the SMEFT for phenomenological applications.
In order to discuss the present and future of such efforts, the ``SMEFT-Tools 2019" Workshop was held at the IPPP Durham on the 12th-14th June 2019. Here we collect and summarize the contents of this workshop.

\setcounter{page}{0}
\thispagestyle{empty}
\newpage

\setcounter{tocdepth}{1}
\tableofcontents

%
%
%

\newpage
\section{Introduction}

Testing the Standard Model (SM) and searching for New Physics (NP) are among the main priorities in High-Energy Physics. Whether or not new particles are directly produced at the LHC, indirect searches will remain crucially important to test the SM and to characterize possible NP patterns.
Indirect searches, defined as searches for far-off-shell effects from  new degrees of freedom, are best framed in the context of Effective Field Theories (EFTs). Constraints from direct searches indicate that these new degrees of freedom, if present, appear at scales much above the Electroweak (EW) scale. Therefore, the relevant EFT for the study of beyond-the-SM (BSM) physics at the EW scale is
the  Standard Model Effective Field Theory (SMEFT)~\cite{Buchmuller:1985jz,Grzadkowski:2010es}. 
For observables at lower energies, such as hadron and lepton decays, other EFTs must be constructed where particles with weak-scale masses have been integrated out~\cite{Buchalla:1995vs,Aebischer:2017gaw,Jenkins:2017jig}. The EFT below the EW scale has been called the Weak Effective Theory (WET) or the Low Energy EFT (LEFT).

Thus, the bulk of any phenomenological analysis of heavy BSM physics at low energies involves a series of steps of matching and renormalization-group evolution, followed by the calculation of low-energy observables.
Since such a procedure is tedious in practice, considerable effort is being devoted to developing public software designed to perform these tasks in an automatic and generic manner. 
Some of the available codes are:

\begin{itemize}

\item Construction of EFT Bases: \texttt{DEFT} \cite{Gripaios:2018zrz}, \texttt{BasisGen} \cite{Criado:2019ugp}, \texttt{abc-eft}~\cite{abceft} .

\item Feynman Rules for the SMEFT: \texttt{SmeftFR} \cite{Dedes:2019uzs}.

\item Matching calculators: \texttt{MatchingTools}~\cite{Criado:2017khh}, \texttt{Matchmaker}~\cite{matchmaker}, \texttt{CoDEx} \cite{Bakshi:2018ics}

\item Generic Matching+Running codes: \texttt{DsixTools} \cite{Celis:2017hod}, \texttt{wilson} \cite{Aebischer:2018bkb}.

\item Fitters: \texttt{SMEFiT}~\cite{Hartland:2019bjb}, \texttt{smelli} \cite{Aebischer:2018iyb}.

\item Observables and Montecarlo enablers: \texttt{EOS} \cite{EOS}, \texttt{flavio} \cite{Straub:2018kue}, \texttt{SMEFTsim} \cite{Brivio:2017btx}.
\end{itemize}

In this context, the 1st Workshop on Tools for Low-Energy SMEFT Phenomenology, {\bf SMEFT-Tools 2019}:
\begin{center}
  \textcolor{blue}{ \href{https://indico.cern.ch/event/787665/}{https://indico.cern.ch/event/787665/}}
\end{center}
was held at IPPP Durham from 12-14th June 2019 with the purpose 
of discussing the status and future prospects of computing tools designed for phenomenological analyses of the SMEFT and the WET/LEFT. This report summarizes the contents of the workshop, complementing the slides available on the web.
We believe that collecting
brief descriptions of most of the currently available tools in this single
document is going to be useful for the community.


\newpage
\talk{BasisGen: counting EFT operators}{Juan Carlos Criado}{University of
  Granada}

\noindent
BasisGen is a Python package that generates bases of operators for EFTs. An EFT is specified by giving its symmetries and field content.
The package takes this information and produces a list of all possible forms for the invariant operators in the theory, together with the number independent operators of each form. It uses well-known methods in representation theory~\cite{Slansky:1981yr}, based on roots and weights, which allow for general and fast calculations\footnote{Other methods include: the Hilbert series~\cite{Lehman:2015via,Henning:2015daa,Lehman:2015coa,Henning:2015alf}, which is similar to BasisGen's approach in generality, organization of the calculation and the structure the results; and the code DEFT~\cite{Gripaios:2018zrz}, which takes a completely different approach.}. To deal with integration by parts redundancy, an adaptation of the method in ref.~\cite{Henning:2015alf} is used.

BasisGen works with any internal symmetry group of the form $G \times U(1)^N$,
where $G$ is a semisimple Lie group. For simplicity, it assumes 4-dimensional
Lorentz invariance. The fields can be in any irreducible representation of both
the internal symmetry group and the Lorentz group. BasisGen generates complete
sets of independent operators, taking into account group-theory identities and
integration by parts. Optionally, redundancies arising from field
redefinitions~\cite{%
  Politzer:1980me,Arzt:1993gz,Criado:2018sdb%
} can be eliminated. For the purpose of generating a basis, this is the same as
using the equations of motion to remove operators~\cite{%
  Georgi:1991ch,Grzadkowski:2003tf,AguilarSaavedra:2009mx%
}.

A module containing the definition of the SMEFT is included in the package. It
can be used to obtain bases of this theory with operators of arbitrary maximum
dimension\footnote{The output of BasisGen for this case has been checked using
  the results in ref.~\cite{Henning:2015alf}.}, but also as a starting point for
defining new theories with extra fields or symmetries. Generating a dimension-6
basis takes only a few seconds in a modern laptop\footnote{3 seconds in a 2.6
  GHz Intel Core i5.}.

Other features of BasisGen include:
\begin{itemize}
\item An interface to its representation-theory functionalities, allowing the
  user to obtain the weight system of any irreducible representation, compute
  tensor products, and decompose reducible representations.
\item The possibility of generating all (not necessarily invariant) operators,
  decomposing their representations into irreducible ones. This gives all the
  possible representations of new fields that couple linearly to the fields in
  the EFT. Such new fields are often the most relevant for
  phenomenology~\cite{deBlas:2017xtg}.
\end{itemize}

A description of the implementation and interface of BasisGen can be found in
ref.~\cite{Criado:2019ugp}. Its code can be downloaded from the GitHub
repository~\url{https://github.com/jccriado/basisgen}. It requires Python
version 3.5 or higher. It can be installed using \texttt{pip} by running:
\begin{minted}{bash}
  > pip install basisgen
\end{minted}

As an example of use, we consider here an EFT with symmetry group
$SU(2) \times U(1)$, whose field content consists of a complex scalar doublet
$\phi$ with hypercharge $1/2$. To import the necessary parts of the package,
once it is installed, one can do
\begin{minted}{python3}
  from basisgen import Field, EFT, scalar, irrep, algebra
\end{minted}
The EFT is defined by
\begin{minted}{python3}
  phi = Field(
      name='phi',
      lorentz_irrep=scalar,
      internal_irrep=irrep('SU2', '1'),
      charges=[1/2]
  )
  my_eft = EFT(algebra('SU2'), [phi, phi.conjugate])
\end{minted}
Now, the following code can be used to compute a basis of operators with maximum
dimension~8:
\begin{minted}{python3}
  invariants = my_eft.invariants(max_dimension=8)
  print(invariants)
  print("Total:", invariants.count())
\end{minted}
The output is:
\begin{minted}{text}
  phi phi*: 1
  (phi)^2 (phi*)^2: 1
  (phi)^2 (phi*)^2 D^2: 2
  (phi)^2 (phi*)^2 D^4: 3
  (phi)^3 (phi*)^3: 1
  (phi)^3 (phi*)^3 D^2: 2
  (phi)^4 (phi*)^4: 1
  Total: 11
\end{minted}
At the beginning of each line, a possible field content for an operator is
given. This is the number of derivatives and fields of each type that the
operator may contain. The positive integer after the colon is the number of
independent invariant operators that can be constructed out of the corresponding
field content.



\newpage
\talk{Automatic Basis Change for Effective Field Theories}{Peter Stangl}{
LAPTh Annecy
}

\noindent
When working with an effective field theory (EFT), it is advisable to use a complete operator basis.\footnote{For a discussion of problems that can arise in ad-hoc phenomenological Lagrangians that do not form a complete basis, see e.g.~\cite{Brivio:2017vri}.
}
Field redefinitions and field relations like equations of motion (EOM) or Fierz identities relate different operators to each other and thus can be used to change the basis.
This is e.g.\ necessary if a matching or loop computation yields operators that are not contained in the desired basis.
Also combining results from analyses performed in different bases or employing renormalization group equations that have only been determined in one specific basis might require a basis change.
While the translation of a small set of operators to a specific basis can be carried out by hand, a full basis change in an EFT that contains hundreds or thousands of operators certainly calls for some automation.
Several codes have been developed to address this problem~\cite{Falkowski:2015wza,Aebischer:2018bkb,Gripaios:2018zrz}.
However, none of the currently available codes is able to perform basis changes of arbitrary bases with the full fermion flavour structure.\footnote{%
\texttt{wilson}~\cite{Aebischer:2018bkb} is able to perform basis changes with the full fermion flavour structure but only for some predefined bases; \texttt{DEFT}~\cite{Gripaios:2018zrz} can perform basis changes of arbitrary bases but only for a single fermion flavour.
}
To close this gap, the new code \texttt{abc\_eft} (\textbf{A}utomatic \textbf{B}asis \textbf{C}hange for \textbf{E}ffective \textbf{F}ield \textbf{T}heories) is currently in development.

The strategy used in \texttt{abc\_eft} to perform automatic basis changes in a given EFT is as follows:
\begin{itemize}
 \item A redundant set of $n_j$ possible operators $O_j$ is generated.
 \item Various linear transformations (Fierz identities, EOMs, integration by part identities, etc.) are applied to each operator.
   All these transformations constitute a matrix $M_{ij}$ of linear dependencies,
   \begin{equation}
    M_{ij}\,O_j = 0.
   \end{equation}
   \item To improve the efficiency of applying numerical matrix algorithms, one can make use of the fact that not all operators are related to each other and therefore $M$ can be turned into a block-diagonal matrix $\widetilde{M} = M\, P^T$.
   $P$ is a permutation matrix that permutes the operators in the vector $O$ in such a way that operators related to each other are grouped together, forming the vector $\widetilde{O}=P\, O$.
Due to the block-diagonality of $\widetilde{M}$, the linear dependencies can be decomposed as
\begin{equation}\label{eq:lin_dep_block}
  M^{(k)}_{lm}\,O^{(k)}_m = 0,
  \quad\quad
  k\in[1,n_k],
\end{equation}
where $M^{(k)}$ are the $n_k$ blocks in $\widetilde{M}$ and each $O^{(k)}$ contains only operators related to each other.
Applying numerical algorithms to the matrices $M^{(k)}$ is much more efficient than applying them to the commonly very large matrix $M$.

\item Each of the independent relations in $M^{(k)}$ can be used to eliminate one of the $n_m^{(k)}$ operators $O^{(k)}_m$ such that the number $n_b^{(k)}$ of basis operators and the number $n_{\bar{b}}^{(k)}$ of non-basis operators among the $O^{(k)}_m$ are given by
\begin{equation}
 n_b^{(k)} = n_m^{(k)} - n_{\bar{b}}^{(k)},
 \quad\quad
 n_{\bar{b}}^{(k)} = \text{rank}\left(M^{(k)}\right).
\end{equation}
After choosing $n_b^{(k)}$ basis operators, the operators $O^{(k)}_m$ can be permuted such that Eq. (\ref{eq:lin_dep_block}) can be written as
 \begin{equation}\label{eq:lin_dep_basis}
 \begin{pmatrix}
  M_{lb}^{(k)} & M_{l\bar{b}}^{(k)}
 \end{pmatrix}
 \begin{pmatrix}
  O^{(k)}_b \\
  O^{(k)}_{\bar{b}} \\
 \end{pmatrix}
=0,
\quad\quad
b\in[1,n_b^{(k)}],
\quad
\bar{b}\in[n_b^{(k)}+1,n_m^{(k)}],
 \end{equation}
where $O^{(k)}_b$ and $O^{(k)}_{\bar{b}}$ contain only basis operators and non-basis operators, respectively.

\item The matrices $M_{lb}^{(k)}$ and $M_{l\bar{b}}^{(k)}$ are generally rank deficient.
A numerical QR decomposition with column pivoting can be used to obtain a permutation matrix $P_{lr}$ that is constructed in such a way that the matrix
\begin{equation}
 \hat{M}^{(k)}_{r\bar{b}} = P^T_{rl}\,M^{(k)}_{l\bar{b}},
 \quad\quad
 r\in\left[1,n_{\bar{b}}^{(k)}\right]
\end{equation}
is a square matrix with full rank (see e.g.~\cite{Anderson:1999:LUG:323215}). Multiplying Eq.~(\ref{eq:lin_dep_basis}) from the left by $P^T_{rl}$ yields
\begin{equation}\label{eq:lin_dep_indep}
 \begin{pmatrix}
  \hat{M}_{rb}^{(k)} & \hat{M}_{r\bar{b}}^{(k)}
 \end{pmatrix}
 \begin{pmatrix}
  O^{(k)}_b \\
  O^{(k)}_{\bar{b}} \\
 \end{pmatrix}
=0,
\quad
\quad
r\in\left[1,n_{\bar{b}}^{(k)}\right],
\end{equation}
which contains only independent relations.

\item Multiplying Eq.~(\ref{eq:lin_dep_indep}) from the left by $\big(\hat{M}^{(k)}\big)^{\!-1}_{\bar{b}r}$ results in
\begin{equation}
 \begin{pmatrix}
  -T_{\bar{b}b}^{(k)} & \mathbbm{1}
 \end{pmatrix}
 \begin{pmatrix}
  O^{(k)}_b \\
  O^{(k)}_{\bar{b}} \\
 \end{pmatrix}
=0,
\quad\quad
  T_{\bar{b}b}^{(k)} = -\big(\hat{M}^{(k)}\big)^{\!-1}_{\bar{b}r}\, \hat{M}_{rb}^{(k)},
\end{equation}
where the minus sign in the definition of $T_{\bar{b}b}^{(k)}$ is introduced for convenience.
The matrix $T_{\bar{b}b}^{(k)}$ can be used to perform the basis change by expressing any non-basis operator in terms of basis operators,
 \begin{equation}
O^{(k)}_{\bar{b}}
=
T_{\bar{b}b}^{(k)}\,O^{(k)}_b.
 \end{equation}
\end{itemize}

An early version of \texttt{abc\_eft} was originally developed for transforming four-fermion operators and was used in the numerical analysis of~\cite{Sannino:2017utc}.
It is capable of
\begin{itemize}
 \item Generating four-fermion operators for an EFT with
 \begin{itemize}
  \item an arbitrary symmetry group,
  \item an arbitrary fermion content in (anti)fundamental and singlet representations of the gauge group factors,
  \item the full flavour structure for an arbitrary number of generations.
 \end{itemize}
 \item Relating operators to each other by
 \begin{itemize}
  \item Fierz identities (including Schouten identities for spinors and identities involving $\sigma_{\mu\nu}\otimes\sigma^{\mu\nu}$),
  \item completeness relations of group generators (e.g.\ the so-called ``colour Fierz'') and Schouten identities for SU(2).
 \end{itemize}
 \item Selecting a basis by general requirements, e.g.\
 \begin{itemize}
  \item group index contraction inside bilinears (which can be used e.g.\ to exclude quark-lepton bilinears),
  \item as few tensor operators as possible,
  \item as few operators containing group generators as possible.
 \end{itemize}
\end{itemize}

The current development aims at generalizing the early version of the code by adding support for operators involving gauge bosons as well as scalars in (anti)fundamental and singlet representations of the gauge group factors.
In the course of this generalization, also many new relations between operators like EOMs and integration by parts identities will be added.
Furthermore, interfaces to other codes are envisaged, e.g.\ a generator of \texttt{WCxf}~\cite{Aebischer:2017ugx} basis files and the possibility to export basis translators to \texttt{wilson}~\cite{Aebischer:2018bkb}.

The source code of \texttt{abc\_eft} will be provided in a public repository at \url{https://github.com/abc-eft/abc_eft}.



\newpage
\talk{Basis construction and translation with \DEFT}{Dave Sutherland}{UC Santa Barbara}

\noindent
\DEFT\footnote{Code available at \url{http://web.physics.ucsb.edu/~dwsuth/DEFT/} and described in \cite{Gripaios:2018zrz}.} takes as input a set of fields and their irreps under a set of $SU(N)$-like symmetries. \emph{In principle}, it can output: a list of all possible operators, to a given order; a list of the redundancies between them; an arbitrary operator basis, and a matrix to convert into and between arbitrary operator bases.

\subsection{Representation of symmetries}

In \DEFT, the transformation of a given field under an $SU(N)$ symmetry is encoded in terms of (anti-)symmetric and traceless combinations of upper and lower fundamental indices. Denoting such indices by latin letters $a,b,\ldots = 1,\ldots,N$, the defining $N$ irrep is written as a single upper index $\phi^a$, the $\bar N$ as a single lower index $\phi_a$, the adjoint $N^2 -1$ as an upper and lower index in a traceless comination $\phi^a_b, \phi^a_b \delta^b_a = 0$, and so forth. Conjugation either raises or lowers each index, e.g.,
\begin{equation}
\phi^a \overset{\text{h.c.}}{\leftrightarrow} \phi^\dagger_a \, .
\end{equation}
There are only three invariant tensors comprising upper and lower fundamental indices: the Kronecker delta, and the upper and lower $N$-index Levi-Civita epsilon tensors
\begin{equation}
\delta^a_b ; \quad \epsilon_{a b \ldots z} ; \quad \epsilon^{a b \ldots z} .
\label{eq:DEFTinvtens}
\end{equation}
$U(1)$ symmetries are encoded by a charge assigned to each field. Both $SU(N)$ and $U(1)$ symmetries can be gauged, meaning that covariant derivatives of the field are understood to contain the corresponding vector potential.

The irreps of the Lorentz group are treated as the irreps of two $SU(2)$ groups (i.e. the canonical dotted and undotted Greek indices), with the understanding that, under conjugation, indices are either dotted or undotted:
\begin{equation}
\psi_\alpha \overset{\text{h.c.}}{\leftrightarrow} \psi^\dagger_{\dot \alpha} \, .
\end{equation}

\subsection{Generation of operators and redundancies between them}

In \DEFT's internal machinery, all operators are expressed as linear combinations of `monomial operators'. A monomial operator is a product of fields and covariant derivatives acting thereon, with zero total $U(1)$ charges, and all $SU(N)$ indices of the fields and derivatives contracted into some product of the invariant tensors (\ref{eq:DEFTinvtens}). A list of all monomial operators is generated combinatorially subject to some criterion (e.g. that the operators have mass dimension~$\leq~d$). An example monomial operator from the one-generation Standard Model (effective field theory) is the dipole operator
\begin{equation}
\delta^{\dot\alpha}_{\dot\gamma} \delta^{\dot\beta}_{\dot\delta} \epsilon^{}_{a b } \epsilon^{d e }_{} \delta^A_B \, {H}^{b }_{} {\bar{Q_L}}^{\dot\gamma }_{e A } {d_R}^{\dot\delta B }_{} {\bar{W}}^{a }_{\dot\alpha \dot\beta d } .
\end{equation}

Having so generated a list of monomial operators, the program enumerates all linear combinations of them that do not contribute to S-matrix elements.\footnote{In the case of the EOM relations, the program generates the linear combination of dimension $d$ operators which have no effect at dimension $d$ order in the S-matrix.} They fall into four categories; we illustrate each with an example for the one generation Standard Model:
\begin{itemize}
\item Fierz relations of the form $\epsilon^{\ldots} \epsilon_{\ldots} = \sum \delta^\cdot_\cdot \delta^\cdot_\cdot \ldots \delta^\cdot_\cdot$, as well as its `raised' and `lowered' $SU(2)$ versions (the Schouten identities)
\begin{align}
 \epsilon^{\dot\alpha \dot\beta }_{} {\bar{H}}^{}_{a } {\bar{e_R}}^{\alpha }_{} D_{\alpha \dot\beta} D_{\beta \dot\alpha} {L_L}^{\beta a }_{}  - \epsilon^{\dot\alpha \dot\beta }_{} {\bar{H}}^{}_{a } {\bar{e_R}}^{\alpha }_{} D_{\beta \dot\beta} D_{\alpha \dot\alpha} {L_L}^{\beta a }_{}  \nonumber \\ 
+ \epsilon^{}_{\alpha \beta } \epsilon^{\gamma \delta }_{} \epsilon^{\dot\alpha \dot\beta }_{} {\bar{H}}^{}_{a } {\bar{e_R}}^{\beta }_{} D_{\delta \dot\beta} D_{\gamma \dot\alpha} {L_L}^{\alpha a }_{} = 0 ;
\end{align}
\item Integration by parts identities
\begin{align}
- \epsilon^{\dot\alpha \dot\beta }_{} D_{\alpha \dot\beta} {\bar{L_L}}^{\dot\gamma }_{b } {L_L}^{\alpha a }_{} {\bar{W}}^{b }_{\dot\alpha \dot\gamma a } - \epsilon^{\dot\alpha \dot\beta }_{} {\bar{L_L}}^{\dot\gamma }_{b } D_{\alpha \dot\beta} {L_L}^{\alpha a }_{} {\bar{W}}^{b }_{\dot\alpha \dot\gamma a } \nonumber \\ 
+ \epsilon^{\dot\alpha \dot\beta }_{} {\bar{L_L}}^{\dot\gamma }_{b } {L_L}^{\alpha a }_{} D_{\alpha \dot\alpha} {\bar{W}}^{b }_{\dot\beta \dot\gamma a } = 0 ;
\end{align}
\item A commutator of covariant derivatives can be replaced by field strengths
\begin{align}
-\frac12 i g^\prime \epsilon^{\alpha \beta }_{} {B}^{}_{\beta \gamma } D_{\alpha \dot\alpha} {\bar{e_R}}^{\gamma }_{} {e_R}^{\dot\alpha }_{} + \frac12 i g^\prime \epsilon^{\dot\alpha \dot\beta }_{} {\bar{B}}^{}_{\dot\beta \dot\gamma } D_{\alpha \dot\alpha} {\bar{e_R}}^{\alpha }_{} {e_R}^{\dot\gamma }_{}  \nonumber \\
- \epsilon^{\alpha \beta }_{} \epsilon^{\dot\alpha \dot\beta }_{} D_{\gamma \dot\beta} D_{\beta \dot\gamma} D_{\alpha \dot\alpha} {\bar{e_R}}^{\gamma }_{} {e_R}^{\dot\gamma }_{} + \epsilon^{\alpha \beta }_{} \epsilon^{\dot\alpha \dot\beta }_{} D_{\beta \dot\gamma} D_{\gamma \dot\beta} D_{\alpha \dot\alpha} {\bar{e_R}}^{\gamma }_{} {e_R}^{\dot\gamma }_{} = 0 ;
\end{align}
\item Equation of motion relations (as well as Bianchi identities)
\begin{align}
-\frac12 \epsilon^{}_{\alpha \beta } \epsilon^{\gamma \delta }_{} \epsilon^{\dot\alpha \dot\beta }_{} \epsilon^{}_{a b } D_{\delta \dot\beta} D_{\gamma \dot\alpha} {H}^{b }_{} {Q_L}^{\beta a A }_{} {\bar{u_R}}^{\alpha }_{A } \nonumber \\
+ y_e^\dagger \epsilon^{}_{\alpha \beta } \epsilon^{}_{\gamma \delta } \epsilon^{}_{a b } {\bar{e_R}}^{\delta }_{} {L_L}^{\gamma b }_{} {Q_L}^{\beta a A }_{} {\bar{u_R}}^{\alpha }_{A } 
+ y_u \epsilon^{}_{\alpha \beta } \epsilon^{}_{\dot\alpha \dot\beta } {\bar{Q_L}}^{\dot\beta }_{a B } {Q_L}^{\beta a A }_{} {\bar{u_R}}^{\alpha }_{A } {u_R}^{\dot\alpha B }_{} \nonumber \\
+ y_d^\dagger \epsilon^{}_{\alpha \beta } \epsilon^{}_{\gamma \delta } \epsilon^{}_{a b } {Q_L}^{\delta b A }_{} {Q_L}^{\beta a B }_{} {\bar{u_R}}^{\alpha }_{B } {\bar{d_R}}^{\gamma }_{A } 
 -2 \lambda \epsilon^{}_{\alpha \beta } \epsilon^{}_{a b } {\bar{H}}^{}_{d } {H}^{b }_{} {H}^{d }_{} {Q_L}^{\beta a A }_{} {\bar{u_R}}^{\alpha }_{A } \nonumber \\ = \text{operators of different dimension, which we ignore} \, .
\end{align}

\end{itemize}

\subsection{Reducing to and converting between non-redundant bases}

Each relation is a vector of Wilson coefficients $\sum_i c_i \op_i = 0$, where $i$ indexes the monomial operators $\op_i$. Collect these $c_i$ as the rows of a matrix, which is then put in reduced row echelon form (RREF). This immediately yields a non-redundant basis $\opB_j$ comprising the monomial operators that correspond to columns of the RREF matrix without a leading entry in any row. The non-trivial components of the RREF matrix can be used to express all monomial operators in terms of those that form the basis:
\begin{equation}
\op_i = R_{ij} \opB_j .
\end{equation}

If the user can input another basis $\opB^\prime_i$ in terms of the monomial operators $\op_i$,
\begin{equation}
\opB^\prime_i = S_{ij} \op_j \, ,
\end{equation}
then the program can convert between the two non-redundant bases (and transitively between any two bases expressed in terms of monomial operators)
\begin{equation}
\opB^\prime_i = S_{ij} R_{jk} \opB_k \, .
\end{equation}

\subsection{Status and future development}

The program correctly reproduces the number of operators in the one-generation Standard Model, as well as theories containing subsets of the Standard Model fields, up to and including dimension 8 (although on a laptop it will take about a day to generate a basis at dimension 8). It contains expressions for one-generation dimension 6 Warsaw and SILH bases in terms of monomial operators, and converts correctly between them.

In the future, it may be useful to:
\begin{itemize}
\item add flavour indices;
\item add expressions convert invariants, such as gamma and Gellmann matrices, into epsilons and deltas and vice versa;
\item add interfaces, such as to FeynRules;
\item generate the EOM relations to higher orders, which are necessary for phenomenology beyond dimension 6 in the SMEFT;
\item tabulate higher irreps of $SU(N)$ --- currently \DEFT has the rules for (anti)-fundamental, adjoint and all-symmetric tensors hard coded in;
\item refactor the code, particularly with an eye to speeding it up and increasing user friendliness.
\end{itemize}



\newpage
\talk{$R_\xi$ gauges in the SMEFT}{Miko{\l}aj Misiak}{University of Warsaw}

\noindent
Practical calculations within the SMEFT require introducing convenient
gauge-fixing terms. Effects of dimension-six operators in the $R_\xi$ gauges
have been studied in Refs.~\cite{Dedes:2017zog, Helset:2018fgq}. Here,
following Ref.~\cite{Misiak:2018gvl}, such an analysis is extended to a wide
class of EFTs with operators of arbitrary dimension. We consider a generic
local EFT arising after decoupling of heavy particles whose masses are of the
order of some scale $\Lambda$, assuming linearly realized gauge symmetry. The
Lagrangian reads
\be
\label{lagr}
\Lu ~=~ \Lu^{(4)} ~+~ \sum_{k=1}^\infty \f{1}{\Lambda^k} \sum_i C_i^{(k+4)} Q_i^{(k+4)}\; ,
\ee
where $\Lu^{(4)}$ is the dimension-four part of $\Lu$, while $Q_i^{(k+4)}$
stand for higher-dimension operators. We are interested in situations when the
scalar fields (treated as real and denoted collectively by $\Phi$) acquire a
non-vanishing VEV $\langle \Phi \rangle = v$ such that $\left|v\right| \ll
\Lambda$. If $v$ is not a gauge singlet, some of the gauge fields
$A_\mu^a$ become massive via the Higgs mechanism.

In the context of $R_\xi$ gauge fixing, one should consider all the operators
that contain bilinear terms in $\varphi = \Phi - v$ and in the gauge fields
$A^a_\mu$.  It can be shown (see Ref.~\cite{Misiak:2018gvl} for details) that
equations of motion allow to bring $\Lu$~(\ref{lagr}) into such a
form that all such bilinear terms are either in the scalar potential or in
\be \label{Lu}
\Lu_{J,K} = -\f14 F^a_{\mu\nu}\, J^{ab}[\Phi]\, F^{b\,\mu\nu} ~+~
\f12 (D_\mu \Phi)_i\, K_{ij}[\Phi]\, (D^\mu \Phi)_j\; .
\ee
The symmetric matrices $J$ and $K$ form a series in $1/\Lambda$ with the
leading ($1/\Lambda^0$) contributions coming from $\Lu^{(4)}$.
To study the bosonic kinetic terms, we set $J$ and $K$ to their expectation
values, i.e.  $J^{ab}[\Phi] \to J^{ab}[v]\equiv J^{ab}$ and $K_{ij}[\Phi] \to
K_{ij}[v]\equiv K_{ij}$. Now $\Lu_{J,K}$ can be written as
\be \label{L}
\Lu_{J,K} = -\f14 A_{\mu\nu}^T \,J \,A^{\mu\nu} ~+~ \f12 (D_\mu \Phi)^T \, K\, (D^\mu \Phi) + \ldots\; ,
\ee
where $A_{\mu\nu}^a \equiv \partial_\mu A^a_\nu - \partial_\nu A^a_\mu$, while
ellipses stand for interactions of three or more fields.

We introduce the $R_\xi$ gauge fixing term as follows:\footnote{
In our notation, the gauge couplings are absorbed into structure constants
$f^{abc}$, and into the generators $T^a$ of the representation in which the
real scalar fields $\Phi$ reside.}
\be \label{Lgf}
\Lu_{GF} = -\f{1}{2\xi}\, {\mathcal G}^a J^{ab} {\mathcal G}^b
%
\hspace{1cm} \mbox{with} \hspace{1cm}
%
{\mathcal G}^a = \partial^\mu A_\mu^a -i\xi  (J^{-1})^{ac} \left[  \varphi^T K T^c v \right]\; .
\ee
The bilinear terms in the sum $\Lu_{J,K} + \Lu_{GF}$ read
\bea
\Lu_{\rm kin,mass} &=& -\f14 A^T_{\mu\nu} J A^{\mu\nu}
+\f12 A_\mu^a\left[ v^T T^a K T^b v \right] A^{b\,\mu}
+\f12 (\partial_\mu \varphi)^T K (\partial^\mu \varphi)\nnb\\
&& -\f{1}{2\xi}(\partial^\mu A_\mu)^T J (\partial^\nu A_\nu) -
       \f{\xi}{2} \left[ \varphi^T K T^a v
         \right](J^{-1})^{ab} \left[ v^T T^b K \varphi \right].\;\; \label{psgmass}
\eea
The last term is the would-be Goldstone boson mass matrix that comes solely from
$\Lu_{GF}$. The physical scalar mass terms (coming from the scalar
potential) are not included in the above equation.

To render the kinetic terms canonical, we redefine the fields as
$\tilde{\varphi}_i  = (K^\f12)_{ij} \varphi_j$ and
$\tilde{A}^a_\mu = (J^\f12)^{ab} A^b_\mu$,
which leads to
\bea
\Lu_{\rm kin,mass} &=& -\f14 \tilde{A}^T_{\mu\nu} \tilde{A}^{\mu\nu}
                  +\f12 \tilde{A}_\mu^T (M^T M) \tilde{A}^{\mu}
                  +\f12 (\partial_\mu \tilde{\varphi})^T (\partial^\mu \tilde{\varphi})\nnb\\
&&                -\f{1}{2\xi}(\partial^\mu \tilde{A}_\mu)^T (\partial^\nu \tilde{A}_\nu)
                  -\f{\xi}{2} \tilde{\varphi}^T (M M^T) \tilde{\varphi}\; , \label{mmt}
\eea
where $M_j^{~b} \equiv [K^\f12 (iT^a) v]_j\, (J^{-\f12})^{ab}$.  If the number
of real scalar fields equals $m$, and the number of gauge bosons equals $n$,
then $M$ is a real $m\times n$ matrix. Its Singular Value Decomposition (SVD)
reads $M = U^T \Sigma V$,
with certain orthogonal matrices $U_{m\times m}$ and $V_{n\times n}$, as well
as a diagonal one $\Sigma_{m\times n}$. Consequently,
$M M^T = U^T (\Sigma \Sigma^T ) U$ and
$M^T M = V^T (\Sigma^T \Sigma ) V$.
Therefore, another redefinition of the fields, namely
$\phi_i  = U_{ij} \tilde{\varphi}_j$ and
$W^a_\mu = V^{ab} \tilde{A}^b_\mu$,
gives the diagonal mass matrices
$m_\phi^2 = \Sigma \Sigma^T$ and
$m_W^2 = \Sigma^T \Sigma$.
The Lagrangian including the gauge fixing term has now the desired form in the
mass-eigenstate basis:
\bea \label{Lufin}
\Lu_{\rm kin,mass} &=& -\f14 W^T_{\mu\nu} W^{\mu\nu}
                       +\f12 W_\mu^T m_W^2 W^{\mu}
                       +\f12 (\partial_\mu \phi)^T (\partial^\mu \phi)\nnb\\
&&                     -\f{1}{2\xi}(\partial^\mu W_\mu)^T (\partial^\nu W_\nu)
                       -\f{\xi}{2} \phi^T m_\phi^2 \phi\; .
\eea

Since our gauge-fixing functionals ${\mathcal G}^a$ in Eq.~\eqref{Lgf} are linear in
the fields, the ghost Lagrangian $\Lu_{FP}$ can be derived from
the Fadeev-Popov determinant. The kinetic terms and interactions for ghosts
$N^a$ and antighosts $\bar N^a$ are then obtained from the variation of
${\mathcal G}^a$ under infinitesimal gauge transformations
$\delta \varphi = -i \alpha^a T^a \left( \varphi + v \right)$~ and~
$\delta A_\mu^a = \partial_\mu \alpha^a - f^{abc} A_\mu^b \alpha^c$.
Taking $\alpha^a(x) = \epsilon N^a(x)$ with an infinitesimal anticommuting
constant $\epsilon$, one gets the BRST
variations
$\delta_{\scs \rm BRST} \varphi =  -i \epsilon N^a T^a \left( \varphi + v \right)$ and
$\delta_{\scs \rm BRST} A_\mu^a =  \epsilon \left( \partial_\mu N^a -  f^{abc} A_\mu^b N^c \right)$,
which determines $M_F^{ab}$~ in~
$\delta_{\scs \rm BRST} {\mathcal G}^a = \epsilon M_F^{ab} N^b$.
The ghost Lagrangian reads
\bea
\Lu_{FP} ~=~ \bar N^a J^{ab} M_F^{bc}\, N^d\nnb &=&
             J^{ab} \bar N^a \raisebox{-.5mm}{$\Box$} N^b
              + \xi \bar N^a [v^T T^a K T^b v] N^b\nnb\\[1mm]
&&\hspace{-1cm} + \bar N^a\, \mbox{${\raisebox{1.2mm}{\boldmath ${}^\leftarrow$}\hspace{-3mm} \partial}$}{}^{\,\mu}
                         J^{ab} f^{bcd} A_\mu^c N^d
              + \xi \bar N^a [v^T T^a K T^b \varphi ] N^b\; . \label{LuFP}
\eea
The BRST variations of ghosts 
take the standard form
$\delta_{\scs \rm BRST} N^a = \frac{\epsilon}{2} f^{abc} N^b N^c$
and
$\delta_{\scs \rm BRST} \bar N^a = \frac{\epsilon}{\xi} {\mathcal G}^a$.
Expressing $\Lu_{FP}$ in terms of ghost mass eigenstates
$\eta = V J^\f12 N$ and $\bar \eta =  V J^\f12 \bar N$,
one finds
$\Lu_{FP} ~=~ \bar \eta^T \raisebox{-.5mm}{$\Box$} \eta ~+~ \xi\, \bar \eta^T m_W^2 \eta ~+~ \mbox{(interactions)}$.

Let us now consider the electroweak sector of SMEFT.  When the Higgs doublet
is written in terms of four real fields $\Phi = (\phi_1, \phi_2, \phi_3,
\phi_4)$, the generators $T^a$ in its covariant derivative $D_\mu \Phi =
\left(\partial_\mu + i T^{a} V^a_\mu\right)\Phi$ can be chosen as
\bea
T^{1} = \f{ig}{2} S \left(\begin{array}{cc} \mbox{\bf 0}_{\scs 2\times 2}  & \sigma^1
  \\ -\sigma^1 & \mbox{\bf 0}_{\scs 2\times 2}  \end{array}\right) S^T\; ,
&
T^{2} = \f{g}{2} S \left(\begin{array}{cc} \sigma^2 & \mbox{\bf 0}_{\scs 2\times 2}  \\
\mbox{\bf 0}_{\scs 2\times 2}  & \sigma^2 \end{array}\right) S^T\; , \nonumber\\[2mm]
T^{3} = \f{ig}{2} S \left(\begin{array}{cc} \mbox{\bf 0}_{\scs 2\times 2}  & \sigma^3
  \\ -\sigma^3 & \mbox{\bf 0}_{\scs 2\times 2}  \end{array}\right) S^T\; ,
&
T^{4} = \f{ig'}{2} S \left(\begin{array}{rrr}
 \mbox{\bf 0}_{\scs 2\times 2} && \mbox{\bf 1}_{\scs 2\times 2}  \\
-\mbox{\bf 1}_{\scs 2\times 2} && \mbox{\bf 0}_{\scs 2\times 2}
\end{array}\right) S^T\; ,
\eea
where $V_\mu^a = (W_\mu^1,W_\mu^2,W_\mu^3,B_\mu)$ and
\be
S = \left(\begin{array}{rrrrrrr}
  0 &\;&  0 &\;&  1 &&  0 \\
  1 &&  0 &&  0 &&  0 \\
  0 &&  0 &&  0 && -1 \\
  0 &&  1 &&  0 &&  0
\end{array}\right).
\ee
The matrices $T^a$ are proportional to those in Eq.~(9) of
Ref.~\cite{Helset:2018fgq}. After the Higgs field takes its VEV $\langle \Phi
\rangle = (0,0,0,v)$, the surviving $U(1)_{\rm em}$
constrains the gauge boson kinetic matrix $J$ to the block-diagonal form
\be
J =  \left(\begin{array}{cc}
   J_C  & \mbox{\bf 0}_{\scs 2\times 2}\\
  \mbox{\bf 0}_{\scs 2\times 2} & J_N \\
\end{array}\right),
\hspace{1cm} \mbox{with} \hspace{1cm}
J_N =  \left(\begin{array}{cc}
   1+J_1  & J_3\\
    J_3  & 1+J_2 \\
\end{array}\right)\; ,
\ee
and $J_C = (1+J_+) \mbox{\bf 1}_{\scs 2\times 2}$. The same argument ensures identical block-diagonal structure of the scalar
kinetic matrix $K$ and, in consequence, of the matrices $M$,
$U$, $V$ and $\Sigma$.
In the charged sector, one finds $\Sigma_C = M_W \, \mbox{\bf 1}_{\scs 2\times
2}$ and $U_C = V_C = \mbox{\bf 1}_{\scs 2\times 2}$, with the charged
$W$-boson mass squared equal to
$M_W^2 = \f14 g^2 v^2 (1 + K_+)/(1 + J_+)$.
In the neutral sector, let us denote $J'_i = 1 + J_i + \sqrt{{\rm det} J_N}$,
for $i=1,2$. Then the matrices appearing in the SVD for the
neutral sector are $\Sigma_N = {\rm diag}(M_Z,0)$,
\be \label{unvn}
U_N = \left(\begin{array}{rrr}
\cos\omega  && \sin\omega \\
-\sin\omega && \cos\omega \\
\end{array}\right)
\hspace{1cm} \mbox{and} \hspace{1cm}
V_N =
\left(\begin{array}{rrr}
\cos\theta && -\sin\theta \\
\sin\theta && \cos\theta
\end{array}\right)\; ,
\ee
where~ $\omega = \arctan(K_3/K'_1)$~ and~ $\theta = \arctan[(g' J'_1 + g
J_3)/(g J'_2 + g' J_3)]$. The $Z$ boson mass squared equals to
\bea
M_Z^2 = \f{v^2}{4} \left( g^2 + g'{}^2 + g'{}^2 J_1 + 2 g g' J_3 + g^2 J_2 \right) \f{1 + K_1}{{\rm det}\, J_N}\; .
\eea
The above expressions hold to all orders in $v/\Lambda$. The leading
effects beyond $\Lu^{(4)}$ arise at ${\mathcal O}(v^2/\Lambda^2)$, in which case one finds
\bea
J_+ = J_1 = -\f{2v^2}{\Lambda^2}C^{\varphi W}\; , \qquad
J_2 = -\f{2v^2}{\Lambda^2}C^{\varphi B}\; , \qquad
J_3 = \f{v^2}{\Lambda^2}C^{\varphi WB}\; ,
\eea
\bea
K_+ = K_3 = 0\; , \qquad
K_1 =  \f{v^2}{2\Lambda^2} C^{\varphi D}\; ,\qquad
K_2 = \f{v^2}{2\Lambda^2} (C^{\varphi D} - 4 C^{\varphi\Box })\; ,
\eea
where $C^{\varphi W}$, \ldots, $C^{\varphi\Box }$ denote the Wilson coefficients
in the Warsaw basis~\cite{Grzadkowski:2010es}. The results of
Ref.~\cite{Dedes:2017zog} can be recovered after introducing the effective
gauge couplings $\bar g = g/\sqrt{1+J_1}$, $\bar{g}' = g'/\sqrt{1+J_2}$, and
then expanding in $v/\Lambda$ up to ${\mathcal O}(v^2/\Lambda^2)$.



\newpage
\talk{The \sfr code}{Athanasios Dedes}{
University of Ioannina
}

\noindent
The \sfr code generates the full set of Feynman Rules (FRs) in linear $R_\xi$-gauges 
for the SMEFT in Warsaw basis.  \sfr is written 
in {\tt Mathematica} language and uses facilities from the \frules program.
This contribution outlines the main features of the code and  is solely based on Refs.~\cite{Dedes:2017zog,Dedes:2019uzs} and  references therein.

 Effective Field Theories (EFTs) are (mostly) useful when certain terms are forbidden 
in a Lagrangian.
As an example, the only \emph{known} problem in the Standard Model (SM) of \emph{electroweak} interactions, 
 that despite observation it predicts massless neutrinos, can easily be addressed by the $d=5$
 Weinberg's  operator leading 
 to Majorana neutrino masses after Electroweak Symmetry Breaking (EWSB)
\begin{equation}
 \frac{C^{\nu\nu}}{\Lambda}\:
(\tilde{\varphi}^\dagger \ell_L)^T  \: \mathbb{C}  \: (\tilde{\varphi}^\dagger \ell_L) \:.
\label{eq1}
\end{equation}
From then on, one can easily construct a renormalizable model by completing 
the SM portals with heavy fields that upon decoupling at scale $\Lambda$
 result in operator (\ref{eq1}).

 For whatever other reason, e.g. dark matter, $(g-2)_\mu$-anomaly, etc., 
 it  could be there is New Physics (NP)  that is related to the SM.  
EFT is then useful to parametrize our ignorance for the size of these effects. 
The parametrization, however, is basis dependent. 
Moreover,  SM is very well measured in gauge sector O(1/200), less in the quark and lepton sectors O(1\%) and far less in the Higgs sector O(15\%). 
Experimental bounds on the relevant Wilson coefficients associated with the new operators have inevitably turned most of  these coefficients
into their perturbative regime so that we can perform higher order corrections as normal.
\sfr code creates all primitive vertices associated with Wilson coefficients in Warsaw basis
while propagators for physical and unphysical fields remain exactly in the same form as
in the Standard Model.

Warsaw basis is written in terms of fields in gauge basis. Following Ref.~\cite{Dedes:2017zog},   \sfr performs field rotations 
and redefinitions to create Feynman Rules in mass basis according to the following steps:

\begin{enumerate}




 \item[A.] We perform a suitable rescaling of gauge fields and  gauge 
couplings $$ \mathcal{L}(W_{\mu\nu}^I, W_\mu^I,...  ; g,g^\prime,  ...) \rightarrow 
\mathcal{L}(\bar{W}_{\mu\nu}^I, \bar{W}_\mu^I,... ; \bar{g}, \bar{g}^\prime,  ...)\ ,$$  such that  gauge kinetic terms become 
canonical after EWSB.
In the end Feynman rules  are written in terms of  the ``barred'' parameters and fields.

\item[B.] Introduce gauge fixing terms such that after EWSB we obtain 
the familiar SM form with gauge fixing parameters $\xi_A, \xi_Z, \xi_W, \xi_G$.

 
\item[C.] Add Faddeev-Popov terms to compensate and restore generalized (BRST) gauge invariance. 

 
\item[D.] Diagonalize mass terms to obtain fields and parameters in mass basis. 
 
 \end{enumerate}
In SMEFT with all $d \le 6$ operators and no expansion in flavour indices,
there are about 120 vertices in unitary gauge and  380 vertices in 
$R_\xi$-gauges.
The structure and the deliverables of \sfr code are synopsized below:
\begin{enumerate}
\item  The SM Lagrangian plus  extra operators in Warsaw basis
  are encoded using {\tt FeynRules} syntax. More specifically:
  \begin{itemize}
  \item {\tt FeynRules} ``model files'' generated dynamically for
    user-chosen subset of operators  
  \item  general flavor structure of all Wilson coefficients
   is  assumed 
   
  \item numerical values of Wilson coefficients (including flavor- and
    CP-violating ones) are  imported from standard files in {\tt WCxf}
    (``Wilson coefficient exchange format'') -- could be interfaced to
    other SMEFT public packages  such as  {{\tt Flavio, FlavorKit,
        Spheno, DSixTools, wilson, FormFlavor, SMEFTSim,
        \dots}}

\item gauge choice user-defined option ({{Unitary}} or
  {{$R_\xi$-gauges}})
  
  \item neutrino masses are incorporated in mass basis
\end{itemize}
\item Derivation of the  SMEFT Lagrangian in mass-eigenstate basis,
  expanded consistently up-to-order {$1/\Lambda^2$}
\end{enumerate}



\begin{enumerate}
\setcounter{enumi}{2}
\item Evaluation of FRs in mass basis, available in several formats useful for further consideration:
  \begin{itemize}
    \item {\tt Mathematica/FeynRules}
    \item {\tt Latex/Axodraw} --  \sfr  here uses a dedicated  generator
    \item {\tt UFO} format -- it can be  imported by ``event generators"
      
      \item {\tt FeynArts} -- it can be imported by  ``symbolic calculators'' 
  \end{itemize}
\item Various options available, such as
  \begin{itemize}
    \item neutrino fields treated as massless Weyl or massive Majorana
      (in the presence of $d=5$ Weinberg operator) spinors
    \item correction of {\tt FeynRules} 4-fermion sign issues
    \item corrected B-, L- violating 4-fermion vertices and 4-$\nu$
      vertex
  \end{itemize}
\end{enumerate}

Here is an example after running \sfr (to see how please consult~\cite{Dedes:2019uzs})\\ 
\begin{center}
\includegraphics[scale=1]{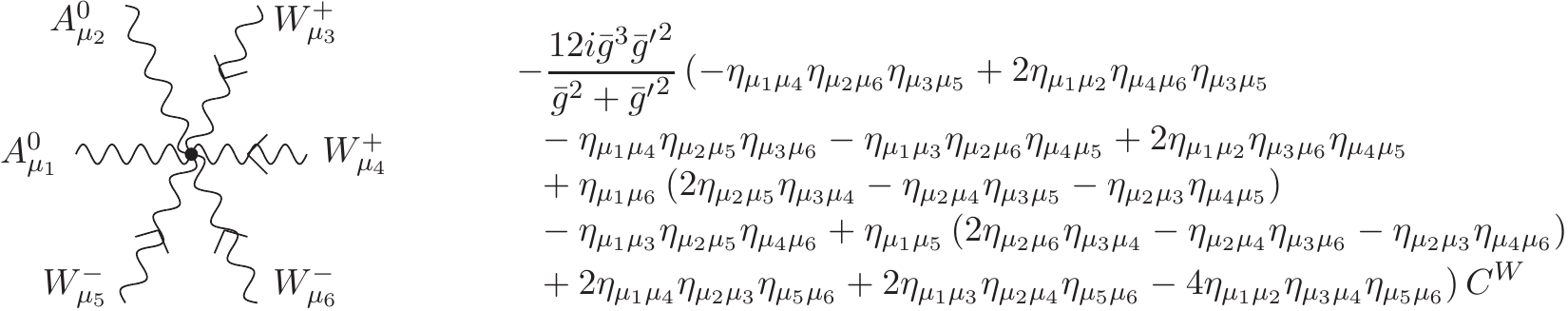} 
\end{center}
\medskip
To the left \sfr draws a (spectacular!)
six leg ``4-$W$-fusion'' vertex into two photons. To the right  \sfr calculates 
the vertex Feynman rule. $C^W$ is the Wilson coefficient associated with the three $W$-field strength operator in Warsaw basis, $\bar{g}, \bar{g}^\prime$ 
are the ``barred'' gauge couplings 
(the perturbative couplings used in SMEFT as we pointed above), and $\eta^{\mu\nu}$ is the Minkowski metric.

We have performed many checks of the generated FRs by \sfr.
For instance, we have checked $\xi$-independence in many tree level amplitudes including 
full flavour structure and lepton number non-conservation, as well as at loop level 
for $\xi$-independence and Ward Identities 
in processes like $h\to \gamma\gamma$ and $h\to Z\gamma$.
We have also made various checks in
 {\tt WCxf} input and output, cross section limits in {\tt UFO/Madgraph5\_aMC@NLO}  (with only one known problem in $C^{\ell e q u (3)}$ that could be fixed in the UFO file),
  and  {\tt FeynArts/FeynCalc} amplitudes for vector boson 
 scattering. The output is obtained for massless Weyl neutrinos due to problems of 
{\tt FeynRules} and interfaces with fermion number violating, higher 
dimensional operators (i.e., problems related to charge conjugated fields and 
ambiguous fermion flow).

The proliferation of primitive vertices in SMEFT demands computer assistance.
 {\tt SmeftFR} is a code for generating Feynman Rules in SMEFT in Warsaw basis. It is 
so far limited to  $d\le 6$ operators.
 {\tt SmeftFR} calculates the FRs in Unitary or in linear $R_\xi$-gauges.
Its output is provided in {\tt Latex, UFO} and {\tt FeynArts} formats.
For a detailed documentation of \sfr the reader is referred to Ref.~\cite{Dedes:2019uzs}  and 
for even more technical details and guidelines to the website:  
\begin{center}
  \textcolor{blue}{ \href{http://www.fuw.edu.pl/smeft}{http://www.fuw.edu.pl/smeft}}
\end{center}
The maintainer of the code is Janusz Rosiek.



\newpage
\talk{EFT below the electroweak scale}{Peter Stoffer}{Physics Department, UC San Diego}

\noindent
The absence of signals of physics beyond the Standard Model (SM) in direct LHC searches suggests that new particles are either very weakly coupled or much heavier than the electroweak scale. In the second scenario, their effects at energies below the scale of new physics can be described by an effective field theory (EFT). Depending on the assumption on the nature of the Higgs particle, this is either the SMEFT~\cite{Buchmuller:1985jz,Grzadkowski:2010es} or HEFT~\cite{Feruglio:1992wf,Grinstein:2007iv}. For processes below the electroweak scale, another EFT should be used, wherein the heavy SM particles, i.e.\ the top quark, the Higgs scalar, and the heavy gauge bosons, are integrated out. This low-energy effective field theory (LEFT) is a gauge theory invariant only under the unbroken SM groups $SU(3)_c \otimes U(1)_\mathrm{em}$, i.e.\ QCD and QED augmented by a complete set of effective operators. It corresponds to the Fermi theory of weak interaction~\cite{Fermi:1934sk}, but by including all operators invariant under the unbroken gauge groups, not only the effects of the SM weak interaction but also of arbitrary heavy physics beyond the SM can be described. This theory has been extensively studied in the context of $B$ physics. The operator basis relevant for $B$-meson decay and mixing has been constructed in~\cite{Aebischer:2017gaw}. The complete LEFT operator basis up to dimension six in the power counting has been derived in~\cite{Jenkins:2017jig}, where also the tree-level matching to the SMEFT above the weak scale was provided. Note that the LEFT is the correct low-energy theory even if the EFT at the high scale is given by HEFT.

The LEFT is defined by
\begin{align}
	\label{eq:LEFTLagrangian}
	\L_\mathrm{LEFT} = \L_\mathrm{QCD+QED} + \L^{(3)}_{\slashed L} + \sum_{d\ge5} \sum_i L_{i}^{(d)} \O_i^{(d)} \, ,
\end{align}
where the QCD and QED Lagrangian is given by
\begin{align}
	\label{eq:qcdqed}
	\L_{\rm QCD + QED} &= - \frac14 G_{\mu \nu}^A G^{A \mu \nu} -\frac14 F_{\mu \nu} F^{\mu\nu} \nn
		&\quad + \theta_{\rm QCD} \frac{g^2}{32 \pi^2} G_{\mu \nu}^A \widetilde G^{A \mu \nu} +  \theta_{\rm QED} \frac{e^2}{32 \pi^2} F_{\mu \nu} \widetilde F^{\mu \nu} \nn
		&\quad + \sum_{\psi=u,d,e,\nu_L}\overline \psi i \slashed{D} \psi   - \left[ \sum_{\psi=u,d,e}  \overline \psi_{Rr} [M_\psi]_{rs} \psi_{Ls} + \text{h.c.} \right] \, .
\end{align}
The additional operators are the Majorana-neutrino mass terms $\L_{\slashed L}^{(3)}$ at dimension three, as well as operators at dimension five and above. At dimension five, there are photonic dipole operators for all the fermions (including a lepton-number-violating neutrino dipole operator) as well as gluonic dipole operators for the up- and down-type quarks. At dimension six, there are the $CP$-even and $CP$-odd three-gluon operators and a large number of four-fermion operators. The entire list of operators up to dimension six can be found in~\cite{Jenkins:2017jig}, including operators that violate baryon and lepton number.

The complete one-loop running and mixing within the LEFT was derived in~\cite{Jenkins:2017dyc}. Within the SMEFT/LEFT framework, the one-loop renormalization-group equations (RGEs) at the high scale~\cite{Jenkins:2013zja,Jenkins:2013wua,Alonso:2013hga}, the tree-level matching~\cite{Jenkins:2017jig}, and the RGEs below the weak scale~\cite{Jenkins:2017dyc} allow one to consistently take into account all leading-logarithm effects and to describe the effects of heavy physics beyond the SM within one unified framework. The RGEs and matching equations have been implemented in several software tools, many of which were presented at this workshop. Consistent EFT analyses at leading-log accuracy that combine constraints from experiments at very different energy scales can be expected to become standard in the near future.

For certain high-precision observables at low energies it is desirable to extend the analysis beyond leading logarithms, e.g.\ in the context of lepton-flavor-violating processes or $CP$-violating dipole moments. Steps in this direction have been done e.g.\ in~\cite{Pruna:2014asa,Crivellin:2017rmk,Panico:2018hal}. Partial results for the matching at the weak scale at one loop were derived in the context of $B$ physics in~\cite{Aebischer:2015fzz,Hurth:2019ula}. The complete one-loop matching between the SMEFT and the LEFT has recently been derived~\cite{Dekens:2019}. It can be used for fixed-order calculations at one-loop accuracy in cases where the logs are not large, and it presents a first step towards a next-to-leading-log analysis within a resummed framework, which, however, will also require the two-loop anomalous dimensions.

At energies as low as the hadronic scale, additional complications appear due to the non-perturbative nature of QCD. In these low-energy processes, one should not work with perturbative quark and gluon degrees of freedom but rather perform either direct non-perturbative calculations of hadronic matrix elements of effective operators or switch to another effective theory in terms of hadronic degrees of freedom, i.e.\ chiral perturbation theory (\chpt{})~\cite{Weinberg:1968de,Gasser:1983yg,Gasser:1984gg}. In~\cite{Dekens:2018pbu}, the matching of semileptonic LEFT operators to \chpt{} has been discussed, which can be obtained within standard \chpt{} augmented by tensor sources~\cite{Cata:2007ns}. Interestingly, through the non-perturbative matching semileptonic tensor operators can contribute to a purely leptonic process like $\mu\to e\gamma$. Constraints on this lepton-flavor-violating process were then used to derive the best bounds on some semileptonic tensor operators.



\newpage
\talk{DsixTools}{Avelino Vicente}{Instituto de F\'{\i}sica Corpuscular (CSIC-Universitat de Val\`{e}ncia)
}

\noindent
\dsix \cite{Celis:2017hod,dsixweb} is a \mathe package for the
matching and renormalization group evolution from the new physics
scale to the scale of low energy observables. \dsix contains numerical
and analytical routines for the handling of Effective Field Theories. Among other features, \dsix allows the user to perform the
full 1-loop Renormalization Group Evolution of the Wilson coefficients
(WCs) of two EFTs, valid at energies above or below the electroweak
scale. In addition, \dsix also includes routines devoted to the
matching to low-energy effective operators of relevance for
phenomenological studies. It can import and export \json and \yaml
files in the {\tt WCxf} exchange format~\cite{Aebischer:2017ugx},
making it easy to link \dsix to other related tools.

\dsixv \cite{Dsix2} is a new version of \dsix that incorporates new
features and updates. Among many improvements, one can clearly
identify four major novelties:

\begin{itemize}
\item {\bf SMEFT-LEFT full integration}
\end{itemize}

\begin{figure}
\centering
\includegraphics[width=0.45\textwidth]{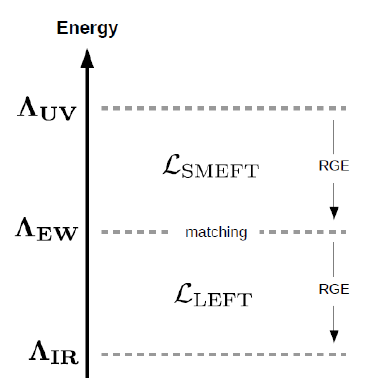}
\caption{\dsixv allows for a complete EFT computation between the high-energy scale $\Lambda_{\rm UV}$ and the low-energy scale $\Lambda_{\rm IR}$.}\label{fig:scheme}
\end{figure}

\dsixv fully integrates two effective field theories: the Standard
Model Effective Field Theory (SMEFT) and the Low-energy Effective
Field Theory (LEFT). While \dsixo placed a special focus on the SMEFT,
\dsixv treats SMEFT and LEFT on an equal footing, including all
operators of both EFTs up to dimension six, their complete 1-loop
Renormalization Group Equations (RGEs) and the tree-level matching
between them. This way, the user can now easily perform a complete
calculation that starts at the high-energy scale $\Lambda_{\rm UV}$,
runs with the SMEFT RGEs down to the electroweak scale, where the
SMEFT is matched onto the LEFT, and then runs with the LEFT RGEs down
to a low-energy scale $\Lambda_{\rm IR}$. This is schematically shown
in Fig.~\ref{fig:scheme}.

\dsix implements the SMEFT in the \textit{Warsaw
  basis}~\cite{Grzadkowski:2010es}. The complete 1-loop RGEs for the
dimension-six operators in this basis have been computed
in~\cite{Jenkins:2013zja,Jenkins:2013wua,Alonso:2013hga,Alonso:2014zka},
whereas the 1-loop RGEs for the dimension-five operators were given
in~\cite{Antusch:2001ck}. For the LEFT, \dsix uses the \textit{San
  Diego basis} introduced in~\cite{Jenkins:2017jig}. The complete
1-loop RGEs for the operators up to dimension six were recently
computed in \cite{Jenkins:2017dyc}. Finally, the tree-level matching
between these two operator bases was given in~\cite{Jenkins:2017jig},
a result that has been independently derived and confirmed as part of
the development of \dsixv.

\newpage

\begin{itemize}
\item {\bf User-friendly input handling}
\end{itemize}

Even after fixing the operator basis for each EFT (Warsaw \& San
Diego), the user must choose a set of operators and make sure that the
input values lead to a consistent Lagrangian. There are two types of
inconsistencies:
\begin{enumerate}
\item {\bf Hermiticity:} The Hermiticity of the Lagrangian imposes certain conditions on some WCs, and these must be respected by the input provided by the user. For instance, an input with $\left( C_{\ell q}^{(1)}\right)_{2223} \ne \left( C_{\ell q}^{(1)}\right)_{2232}^\ast$ would be inconsistent.
\item {\bf Antisimmetry:} Some LEFT operators are antisymmetric under the exchange of two flavor indices and this implies that some WCs must be vanish. For instance, an input with $\left( L_{\nu \gamma} \right)_{11} \ne 0$ would be inconsistent.
\end{enumerate}

In order to avoid potential issues associated to inconsistent inputs,
\dsixv includes user-friendly input routines that simplify the user's
task. The new version of \dsix accepts input values for the WCs of any
set of operators (belonging to the Warsaw or San Diego bases) and then
checks for possible consistency problems. In the event of an
inconsistency, \dsix applies a change in the input to fix it and
displays a warning message to inform the user. Furthermore, \dsix
transforms all Wilson coefficients to the \textit{symmetric basis},
defined as the basis in which the WCs follow the same symmetry
conditions as the associated operators. For instance, in this basis
$\left( C_{\ell \ell} \right)_{1122} = \left( C_{\ell \ell}
\right)_{2211}$ since $\left( Q_{\ell \ell} \right)_{1122} = \left(
Q_{\ell \ell} \right)_{2211}$. This is the basis used internally by
\dsix. Nevertheless, the user needs not to worry about this, since the
input/output is always unambiguous.

\begin{itemize}
\item {\bf More visual and easy to use}
\end{itemize}

\dsixv aims at a simpler and more visual experience. Many changes and
simplifications of the package have been applied in order to guarantee
this. In this new version, the user will be able to obtain the same
results with much shorter programs than before. For instance, a
complete program with a full EFT computation between the high-energy
scale $\Lambda_{\rm UV}$ and the low-energy scale $\Lambda_{\rm IR}$
can now be written with only 6 lines of code. This is possible thanks
to new routines that include automatic multi-step
calculations. Moreover, a new {\tt Dictionary} routine is
available. This routine displays a large amount of useful information
on any WC or operator of the SMEFT or the LEFT specified by the
user. Along the same lines, a more intuitive naming for the SMEFT WCs
is used in \dsixv and more informative error messages are
displayed. Finally, \dsixv also contains an improved documentation. In
addition to a printable manual, a comprehensive documentation system,
fully integrated in \mathe, is also available for the user right after
installing the package.

\begin{itemize}
\item {\bf Evolution matrix formalism}
\end{itemize}

Last but not least, \dsixv implements a new and much faster method to
solve the SMEFT and LEFT RGEs. This new approach is based on an
evolution matrix formalism.

In order to understand the new method one can consider the case of the
SMEFT, the application to the LEFT being analogous. The SMEFT RGEs can
be generically written as (with $t\equiv\ln\mu$)
\begin{align}
\frac{d \hat C_i(t)}{dt}&=\frac{1}{16\pi^2}\,\hat \gamma_{ij}(\hat C_k,C_k)\,\hat C_j(t)\,,\\[5pt]
\label{eq:d6RGEfull}
\frac{d C_i(t)}{dt}&=\frac{1}{16\pi^2}\,\gamma_{ij}(\hat C_k)\, C_j(t)\,, 
\end{align}
where $\gamma$ is the anomalous dimensions matrix (ADM). Quantities
associated to dimension-four ($d=4$) objects are denoted with a hat,
while the non-hatted quantities correspond to the dimension-five and
-six ($d>4$) ones. These form a system of coupled differential
equations. However, one must note that, at first order in the EFT
expansion, the ADM for the $d>4$ operators only depend on $\hat
C_k$. Therefore, since $C_k\sim1/\Lambda_{\rm UV}^2$, the first
equation above can be written as
\begin{align}
\frac{d \hat C_i(t)}{dt}=\frac{1}{16\pi^2}\,\hat \gamma_{ij}(\hat C_k)\,\hat C_j(t)+\mathcal{O}(1/\Lambda_{\rm UV}^2)\,,
\end{align}
which corresponds to the Standard Model RGE evolution. These equations
are known up to 3-loops and can be solved numerically relatively fast,
since they only involve the $d=4$ sub-block, leading to
\begin{align}
\hat C_k(t)=C_k^{\rm SM}(t)+\mathcal{O}(1/\Lambda_{\rm UV}^2)\,,
\end{align}
where $C_k^{\rm SM}(t)$ are interpolating functions obtained by the
numerical RGE solution. One can now plug this into Eq.
\eqref{eq:d6RGEfull} to write
\begin{align}
\frac{d C_i(t)}{dt}=\frac{1}{16\pi^2}\,\gamma_{ij}(C_k^{\rm SM})\, C_j(t)+\mathcal{O}(1/\Lambda_{\rm UV}^2) \equiv \bar \gamma_{ij}(t)\, C_j(t)+\mathcal{O}(1/\Lambda_{\rm UV}^2)\,,
\end{align}
so that the ADM is now a function of $t$ only. Therefore, the
resulting equation can be solved in terms of an evolution matrix $U
\sim \exp \left( \bar \gamma \right)$, such that
\begin{align}
C_i(t)=U_{ij}(t,t_0)\,C_j(t_0)\,.
\end{align}
Finally, once the $C_i(t)$ solutions have been found, they can be
plugged into the equations for the $d=4$ parameters. Given the small
number of $d=4$ operators, these can be solved numerically quite fast,
obtaining in this way a full $\mathcal{O}(1/\Lambda_{\rm UV}^2)$
solution of the system.



\newpage
\talk{Wilson/WCxf}{Speaker Jacky Kumar}{ University of Montreal
}
\label{sec:wilson}

\begin{figure}[H]
\centering
	\hspace{-2cm}
\includegraphics[clip, trim=0.1cm 8cm 0.1cm 8cm,width=1.0\textwidth]{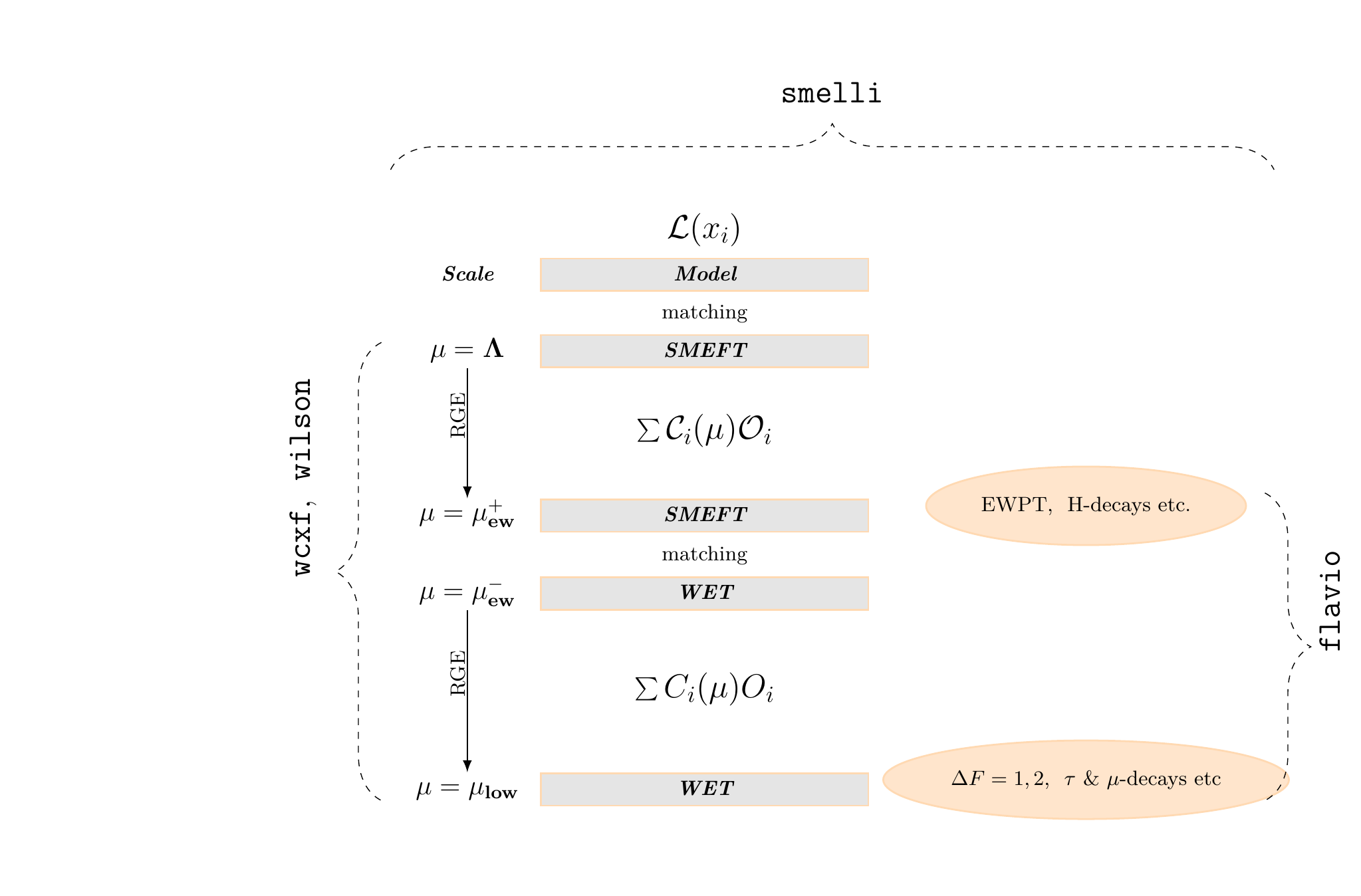}
\caption{The standard steps involved in a phenomenological analysis using SMEFT and
	packages to perform these steps.}
\label{fig:flow}
\end{figure}

\noindent
To study the low energy phenomenology in a model using SMEFT, the first step is to match the model
at the high scale $\Lambda$ to the effective operators $\mathcal{O}_i$ of SMEFT, then to run down the
SMEFT operators to the electroweak scale, match them onto the WET operators $O_i$ and then further run them down
to the mass of the bottom quark or some other low scale depending on the process that we are interested in.
Once we have the Wilson coefficients at a given scale the next step is to calculate the observables
of interest in terms of these Wilson coefficients. These steps are visualized in Fig.\ref{fig:flow}.
The aim of the python program {\tt wilson}\cite{Aebischer:2018bkb} is to automate the running, matching and translating
the bases of Wilson coefficients in SMEFT and WET. The {\tt wilson} package is built upon the
\emph{Wilson coefficient exchange format}({\tt wcxf})\cite{Aebischer:2017ugx}.
It can be used together with packages like {\tt flavio}\cite{Straub:2018kue} for predictions of observables, {\tt smelli}\cite{Aebischer:2018iyb} for global fits\footnote{For more information about {\tt flavio} and {\tt smelli} see the talk by David Straub.} to carryout a
complete phenomenological analysis using SMEFT, or for clustering analyses using the Python package {\tt ClusterKinG} \cite{Aebischer:2019zoe}.

{\tt wcxf} is a data exchange format for the numerical values of Wilson coefficients
which helps interfacing different packages used in particle physics.
For example \emph{model-specific} Wilson coefficient calculators, renormalization group (RG) runners,
and observable calculators etc.

\noindent
The important features of {\tt wcxf} are:
\begin{itemize}
\item  Unambiguous:  it uses a non-redundant set of operators and a fixed
	normalization for each basis.
\item Extensible: it allows the addition of new EFTs and bases by the user.
\item Robust: it is based on {\tt yaml} and {\tt JSON} files.
\end{itemize}
The format is defined in terms of three kinds of files:
\begin{itemize}
\item  The EFT file, which is immutable, fixing the theory such as SMEFT or WET.
\item The basis file, which is also immutable, defining the basis by listing all non-vanishing operators. Examples include the Warsaw or flavio basis.
\item The Wilson coefficient file, which contains the actual data i.e the numerical values of Wilson coefficients
	at a given scale for an EFT in some basis.
\end{itemize}
In the current version of the program, the SMEFT and WET, WET-4, WET-3, WET-2 are implemented.
Here the various WETs differ in the number of quark and lepton flavours.
Furthermore, for the SMEFT the Warsaw \cite{Grzadkowski:2010es}, Warsaw up and
Warsaw mass\cite{Aebischer:2015fzz} bases and for WET the JMS, Bern, flavio, formflavor, FlavorKit and
EOS bases are predefined.

\bigskip
\noindent
The {\tt wilson} package uses the Wilson coefficient exchange format. Given the numerical values of the Wilson coefficients at a given scale it can perform:
\begin{itemize}
\item Running of the complete set of dimension six SMEFT operators \\(based on DsixTools \cite{Celis:2017hod}).
\item Matching (tree level) of SMEFT onto WET for all operators.
\item Running of the complete set of dimension six WET operators.
\item Translation of bases in SMEFT and WET.
\end{itemize}
The implementation of {\tt wilson} based on the following work:
\begin{itemize}
\item The 1-loop SMEFT RGEs are based on anomalous dimension calculations of Refs\cite{Alonso:2013hga, Jenkins:2013zja, Jenkins:2013wua}.
\item The tree level SMEFT onto WET matching is based on the calculations performed in Refs\cite{Aebischer:2015fzz, Jenkins:2017jig}.
\item The 1-loop (QCD, QED) WET running is based on Refs\cite{Aebischer:2017gaw, Jenkins:2017dyc}.
\end{itemize}

To install {\tt wilson} and {\tt wcxf}, one has to execute the
following commands in the terminal:
\begin{center}
{\tt python3 -m pip install wilson {\textit --}user}

{\tt python3 -m pip install wcxf {\textit --}user}
\end{center}

For further information and updates about {\tt wilson} and {\tt wcxf} we refer to the project websites \url{https://wilson-eft.github.io/} and \url{https://wcxf.github.io/} respectively.



\newpage
\talk{The SMEFTsim package}{Ilaria Brivio}{Institut f\"ur Theoretische Physik, Universit\"at Heidelberg 
}

\subsection{Motivation and scope}
The \SMEFTsim\ package~\cite{Brivio:2017btx} is mainly designed for enabling global SMEFT analyses at the LHC. 
It is available at \href{https://feynrules.irmp.ucl.ac.be/wiki/SMEFT}{feynrules.irmp.ucl.ac.be/wiki/SMEFT}.
The users are encouraged to check periodically this repository for updates.
\vskip.5em

\SMEFTsim\ is a  \Feynrules\ and UFO implementation of the full set of  dimension-6, baryon number conserving operators of the Warsaw basis~\cite{Grzadkowski:2010es}. It allows the parton-level Monte Carlo simulation of arbitrary processes in the presence of SMEFT operators.

Its main scope is the estimation of leading SMEFT corrections to SM observables: it is not equipped for NLO evaluations and the effective Lagrangian is truncated at order $\Lambda^{-2}$, corresponding to a numerical accuracy of a few \% for $C_i\sim 1$ and $\Lambda\gtrsim 1$~TeV. Note that, for this reason, 
complete and
fully consistent results are only ensured for $\mathcal{O}(\Lambda^{-2})$ contributions. 

\subsection{Implementation}
The \SMEFTsim\ package contains the \Feynrules\ input files and a set of pre-exported UFO models. The latter have been optimized for \MG
\ but are compatible with most Monte Carlo generators.

The SMEFT Lagrangian is implemented in the fermion basis where the $d$ quark masses are diagonal. The gauge fields are rescaled to bring their kinetic terms to  canonical form and the SM parameters are automatically redefined to account for SMEFT corrections to the input measurements.
The Lagrangian and analytic Feynman rules can be accessed with the \texttt{Mathematica} notebook supplied. The model is \emph{not} equipped for NLO simulations  and gauge choices other than the unitary gauge are not fully supported.

In order to reproduce all the main Higgs production and decay channels in the SM, the loop-induced Higgs couplings ($hGG,\, h\gamma\gamma,\, hZ\gamma$) are implemented as effective vertices with couplings given by the SM $t$- and $W$-loops evaluated for on-shell external bosons. 

Interaction orders are defined in order to control the interactions to be included in the Monte Carlo generation. In addition to the customary \texttt{QCD} and \texttt{QED} orders, all SMEFT vertices have an interaction order \texttt{NP=1}, and the SM loop-induced Higgs couplings have order \texttt{SMHLOOP=1}. In \MG\ one can evaluate e.g. the pure SM-SMEFT interference terms with the syntax \texttt{NP\^{}2 == 1}.

\SMEFTsim\ supports the \wcxf\ exchange format~\cite{Aebischer:2017ugx} via a python script that converts a \wcxf\ input file into a \texttt{param\_card} for the UFO models.

\subsubsection*{Implemented frameworks}
\SMEFTsim\ comes in 6 different implementations, that differ in flavor assumptions (3 options) and input parameter scheme (2 options).

The \underline{flavor assumptions} currently available are the following:
\begin{itemize}
\setlength{\itemsep}{0em}
\item[\tt general] the most general structure, where all flavor indices are free. This setup contains 2499 SMEFT parameters.
\item[\tt U35] a $U(3)$ symmetry is assumed for each of the 5 fermion fields of the SM ($q,l,u,d,e$), resulting in a $U(3)^5$-symmetric Lagrangian. Yukawa couplings are taken to be spurions of this symmetry, and only terms with up to 1 Yukawa insertion are retained. This scenario is the most restrictive and contains 81 SMEFT parameters.
\item[\tt MFV] a linear Minimal Flavor Violation ansatz: the CP and $U(3)^5$ symmetries of the Lagrangian are assumed to be violated only due the breaking sources of the SM, i.e. the CKM complex phase and the Yukawa couplings.
CP violation in purely bosonic operators are suppressed proportional to the Jarlskog invariant $J\sim 3\cdot 10^{-5}$ and are therefore neglected. 
Flavor symmetric spurion insertions up to $\mathcal{O}(y_b^2,y_t^2)$ are retained, in contrast with the {\tt U35} setup where only the leading, Yukawa-indepedent terms are included. This models contains 129 independent parameters.
\end{itemize}
The \underline{input parameters sets} implemented are
$\{\alpha_{em},m_Z,G_F\}$ (which is labeled \texttt{alphaScheme})
and $\{m_W,m_Z,G_F\}$ (labeled \texttt{MwScheme}),
where $m_{W,Z}$ are the masses of the electroweak bosons, $G_F$ is the Fermi constant and $\alpha_{em}$ the fine structure constant. These two sets provide alternative choices for fixing the numerical values of the 3 free parameters of the electroweak gauge sector of the SM (that can be chosen to be e.g. $\{g,g',v\}$). In addition to these, the Higgs mass is used to fix the remaining free parameter in the scalar potential and the fermions' masses to fix the Yukawa couplings (an input scheme for the CKM matrix can also be defined~\cite{Descotes-Genon:2018foz}, although this is not currently implemented in \SMEFTsim). As the input measurements are generically affected by dimension 6 operators, the SM parameters inferred from them are correspondingly shifted, so that a net discrepancy between an input and a predicted measurement is formally moved to the latter (see e.g. Refs.~\cite{Brivio:2017bnu,Brivio:2017btx} for further details). Such parameter shifts are automatically included in the \SMEFTsim\ Lagrangian.

Restriction cards are provided within each UFO model, that can be used e.g. to set to zero light fermion masses or to recover the SM limit. Further ad hoc restrictions can be easily added by the user.

\subsection{Validation}
\SMEFTsim\ has been validated in several ways. Most notably:
\begin{itemize}
\setlength{\itemsep}{0em}
\item with an internal validation. Two independent complete versions (set A and set B) were created, whose output was compared for a large number of processes. Both sets are available online for cross-checks.
\item against \dimsixtop\ for the subset of operators containing a top quark. All the tables in Ref.~\cite{AguilarSaavedra:2018nen} were derived independently with the two UFO models.
\item against both \dimsixtop\ and \SMEFTatNLO\, using the validation tools and procedure recommended by the LHC Top and Electroweak Working Groups and by the LHC Higgs Cross Section Working Group~\cite{Durieux:2019lnv}.
\item against a number of expressions derived with independent analytic calculations, in particular for EW and Higgs observables at LEP and at the LHC (e.g. those in~\cite{Berthier:2015oma,Brivio:2019myy}).
\end{itemize}



\newpage
\talk{The SMEFiT fitting code}{Emma Slade}{Rudolf Peierls Centre for Theoretical Physics, University of Oxford}

\noindent
The effects of dimension-6  operators  in the SMEFT can be written as:
\begin{equation}
\label{eq:smeftXsecInt}
\sigma=\sigma_{\rm SM} + \sum_i^{N_{d6}}
\kappa_i \frac{c_i}{\Lambda^2} +
\sum_{i,j}^{N_{d6}}  \widetilde{\kappa}_{ij} \frac{c_ic_j}{\Lambda^4}  \, ,
\end{equation}
where $\sigma_{\rm SM}$ indicates
the SM prediction and $c_i$ are the Wilson coefficients we wish to constrain.
In this work we develop a novel
strategy for global SMEFT analyses~\cite{Hartland:2019bjb}, which we denote by SMEFiT.
 As a proof of concept of the SMEFiT methodology, we apply it here to the study of top quark production at the LHC in the SMEFT 
framework at dimension-6.
We adopt the Minimal
Flavour Violation (MFV) hypothesis~\cite{DAmbrosio:2002vsn} in the quark
sector as the baseline scenario, impose a $U(2)_q\times U(2)_u \times U(2)_d$ flavour symmetry in the first
two generations and consider CP-even operators, ending up with 34 degrees of freedom.

We adopt the Monte Carlo (MC) replica method as it does not make any 
assumption about the probability distribution of the coefficients, 
and is not limited to Gaussian distributions.
Given an experimental measurement of a
cross-section, denoted by $\mathcal{O}_i^{\rm (exp)}$, with
total uncorrelated uncertainty $\sigma_{i}^{\rm (stat)}$, $N_{\rm sys}$ 
correlated systematic uncertainties $\sigma^{\rm (sys)}_{i,\alpha}$, and
$N_{\rm norm}$ normalisation uncertainties $\sigma^{\rm (norm)}_{i,n}$, the artificial replicas are generated as
\begin{equation}
\label{eq:replicas}
\mathcal{O}_{i}^{(\mathrm{art})(k)}
=
S_{i,N}^{(k)} 
\mathcal{O}_{i}^{\mathrm (\mathrm{exp})}\bigg( 1
+
r_{i}^{(k)}\sigma_{i}^{\rm (stat)}
+
\sum_{\alpha=1}^{N_{\rm sys}}r_{i,\alpha}^{(k)}\sigma^{\rm (sys)}_{i,\alpha}\bigg)
\ , \quad k=1,\ldots,N_{\mathrm{rep}} \ , 
\end{equation}
where the index $i$ runs from 1 to $N_{\rm dat}$ and
where $S_{i,N}^{(k)}$ is a normalisation term.
In order to ensure that no residual MC fluctuations remain, we will use 
$N_{\rm rep}= 1000$ as our baseline.
For each MC replica, the corresponding best-fit values are determined from
the minimisation of a figure of merit
\begin{equation}
  E(\{c_l^{(k)}\})\equiv \frac{1}{N_{\rm dat}}\sum_{i,j=1}^{N_{\rm dat}}\left(
  \mathcal{O}^{(\rm th)}_i\left( \{c_l^{(k)}\} \right)-\mathcal{O}^{{(\rm art)}(k)}_i\right) ({\rm cov}^{-1})_{ij}
  \left( \mathcal{O}^{(\rm th)}_j\left( \{c_l^{(k)}\} \right)-\mathcal{O}^{{(\rm art)}(k)}_j\right)
  \label{eq:chi2definition}
    \; ,
\end{equation}
where $\mathcal{O}^{(\rm th)}_i$ is the theoretical
prediction for the $i-$th cross-section evaluated using the $\{ c_l^{(k)}\}$  
values for the Wilson coefficients.
The final fit 
quality can be quantified with the $\chi^2$
\begin{equation}
  \chi^2 \equiv \frac{1}{N_{\rm dat}}\sum_{i,j=1}^{N_{\rm dat}}\left( 
  \mathcal{O}^{(\rm th)}_i\left( \{ \langle c_l\rangle \} \right)
  -\mathcal{O}^{(\rm exp)}_i\right) ({\rm cov}^{-1})_{ij}
\left( 
  \mathcal{O}^{(\rm th)}_j\left( \{ \langle c_l\rangle \} \right)
  -\mathcal{O}^{(\rm exp)}_j\right)
 \label{eq:chi2definition2}
    \; ,
\end{equation}
where now the theoretical predictions, computed using the expectation value 
 for the degree of freedom $c_l$, are compared to the central 
experimental data.
This is evaluated as the average over the resulting MC best-fit sample 
$\{c_l^{(k)}\}$.

We use as input to all our theory calculations the NNPDF3.1 NNLO
no-top PDF set~\cite{Ball:2017nwa}, to prevent us double-counting the data both in the PDFs and the SMEFT fits. To account for the removal of the data in the PDF fit, we include PDF uncertainties in the  covariance matrix. We use NNLO QCD predictions for all available SM processes, and NLO otherwise.
We also use MC cross-validation to prevent over-fitting the coefficients, and implement closure tests to ensure a rigourous test of the SMEFiT methodology.
%

\begin{figure}
  \begin{center}
\includegraphics[scale=0.5]{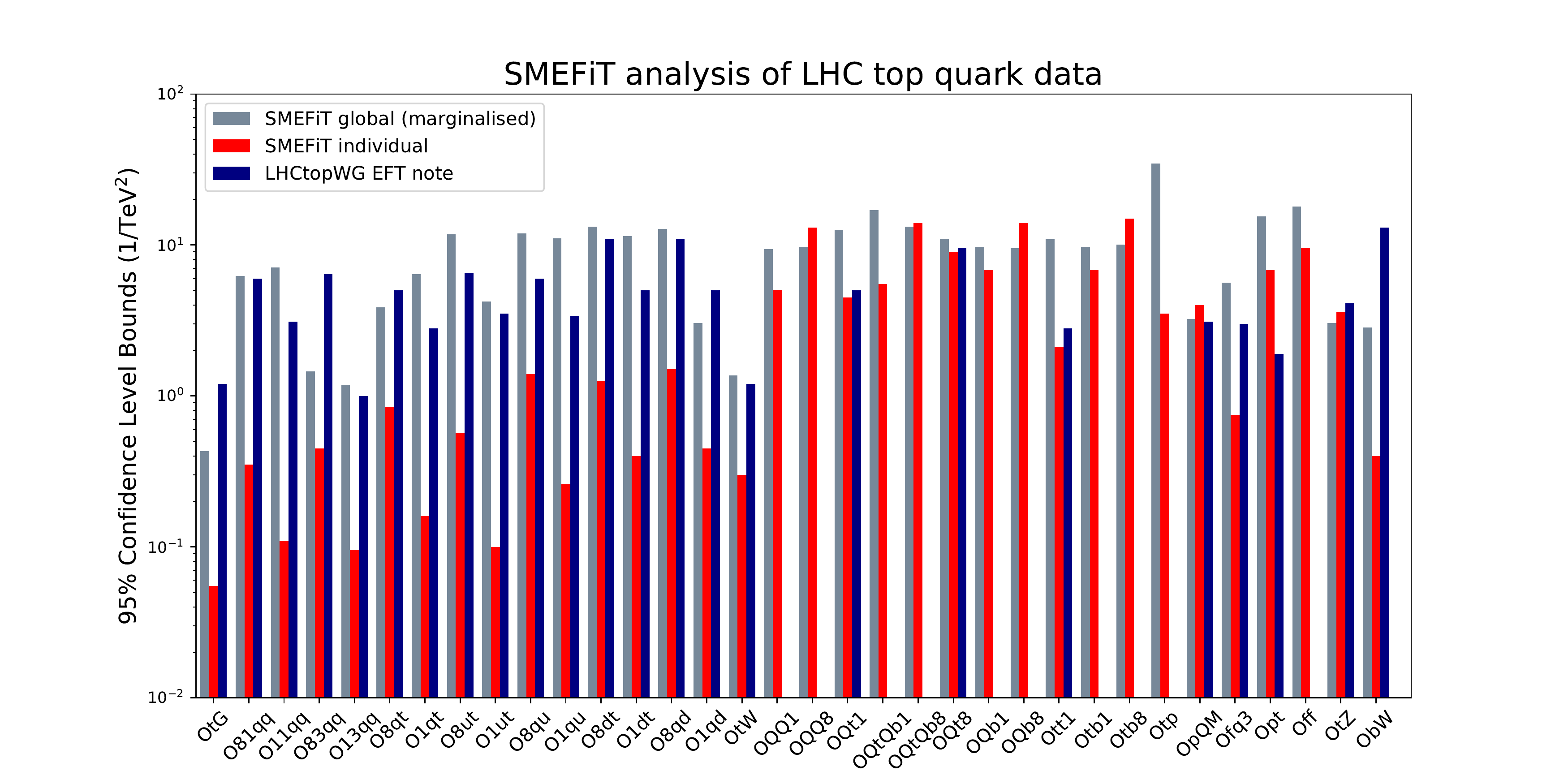}
    \caption{\small The 95\% CL bounds on the 34 degrees of freedom
      included in SMEFiT, both in the marginalised
      and in the individual fit cases, with the
      bounds reported in the LHC Top WG EFT note~\cite{AguilarSaavedra:2018nen}.
The definitions of the  degrees of freedom  is given in Ref.~\cite{Hartland:2019bjb}.
      \label{fig:SMEFiT-bounds}
  }
  \end{center}
\end{figure}

\begin{figure}
\centering
\includegraphics[scale=0.5]{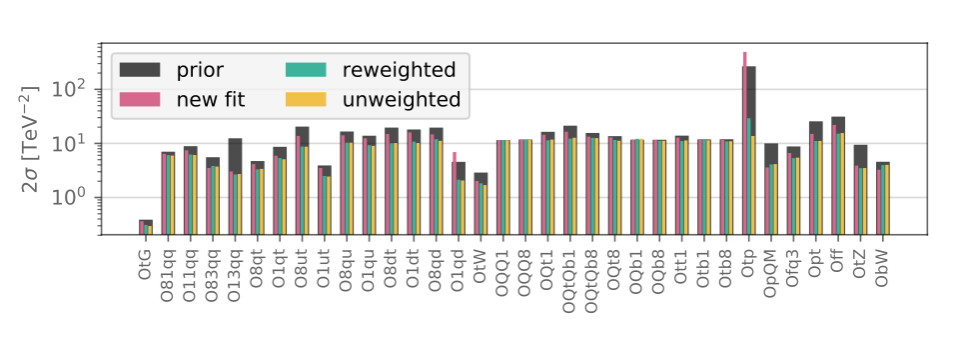}
\caption{\small The 95\% CL bounds for
    the $N_{\rm op}=34$ Wilson coefficients considered in the SMEFiT reweighting analysis 
    of the top quark sector.    %
\label{fig:two_sigma_bounds_t_channel} }
\end{figure}

In Fig.~\ref{fig:SMEFiT-bounds} we show the 95\% CL bounds on the 34  degrees of freedom considered in~\cite{Hartland:2019bjb} both at the marginalised and individual-fit level.
Within finite-size uncertainties, we find, as expected, the individual bounds to be tighter than the global bounds, as correlations between degrees of freedom  are ignored in the former.
Some of us have reported~\cite{vanBeek:2019evb} on the applicability of the Bayesian reweighting technique developed for fitting Parton Distribution Functions~\cite{Ball:2010gb,Ball:2011gg}\footnote{Code may be found at the url \url{https://github.com/juanrojochacon/SMEFiT}}. 
This method has two advantages in comparison to a fit to new data: it is essentially instantaneous, and it can be carried out without access to the SMEFT fitting code.
We show in Fig.~\ref{fig:two_sigma_bounds_t_channel}  the prior results without any single-top 
    data included with those after $t$-channel measurements 
    have been added either by reweighting or by performing a new fit. 
We find that, under well-defined conditions, the results obtained with reweighting all the single-top $t$-channel data are equivalent to those obtained with a new fit to the extended set of data.



\newpage
\talk{Flavour Physics with EOS}{Danny van Dyk}{Technische Universit\"at M\"unchen}

\noindent
Within the Standard Model (SM) of particle physics, changes of flavour quantum numbers follow stringent rules.
Tree-level processes can change flavour only in charged-current processes, while neutral-current processes
emerge first at the one-loop level and are thus suppressed.
The coupling strengths of both types of processes are governed exclusively by the Yukawa couplings.
Measurements of flavour-changing processes therefore provide means to determine most of the Standard Model (SM) parameters,
but also place stringent constraints on possible effects beyond the Standard Model (SM) of particle physics \cite{Isidori:2010kg}.
While commonly used to bound the model parameters of UV-complete theories that could replace the SM, constraints from flavour observables
can also be used to distinguish between different models for the origin of flavour, and between different dynamics
behind electroweak symmetry breaking \cite{Cata:2015lta}.
With the absence of any direct hints of effects Beyond the SM (BSM), indirect flavour constraints have stirred increasing interest
amongst model builders. It is therefore important to the flavour physics community to provide convenient and
easy bridges toward using our results to the any and all interested parties. \EOS is meant to be such a bridge.\\

The \EOS software has been authored with three use cases in mind:
\begin{enumerate}
    \item To produce accurate and precise theory predictions and uncertainty estimates of flavour observables
        and related theoretical quantities. \EOS aspires to facilitate the production of these predictions and estimates
        with publication quality.
    \item To infer a variety of parameters from experimental measurements and from theoretical constraints.
        For this task, \EOS defaults to using a Bayesian statistical framework. For the convenience of the user, \EOS
        ships with a database of measurements and further constraints that can be used immediately.
    \item To produce pseudo events that are useful to carry out sensitivity studies for phenomenological and
        experimental analyses. \EOS aspired to produce such pseudo events for direct use in experimental analyses.
\end{enumerate}
To achieve the outcomes of these use cases as effectively as possible, \EOS has been written as a C++14 software
with Python3 bindings. It includes sophisticated Monte Carlo tools based on Markov chains and importance sampling \cite{Beaujean:2012uj}
to tackle statistical analyses with $\mathcal{O}(100)$ parameters.
Binary packages of \EOS are available for a variety of Linux distributions, and \EOS can
be quickly installed on MacOS via the Homebrew package manager \cite{Homebrew}.\\

\begin{figure}
\centering
\includegraphics[width=.49\textwidth]{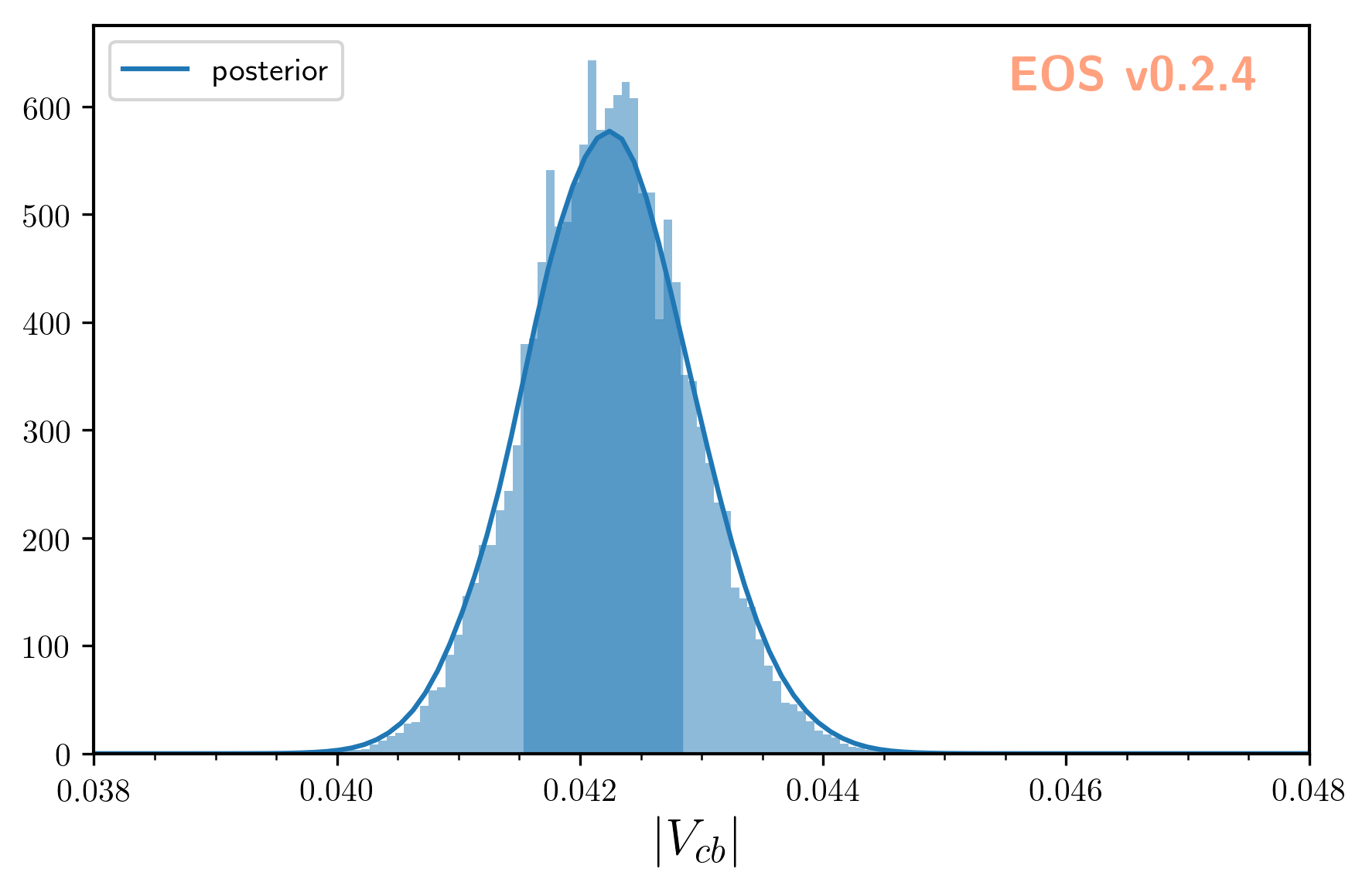}
\includegraphics[width=.49\textwidth]{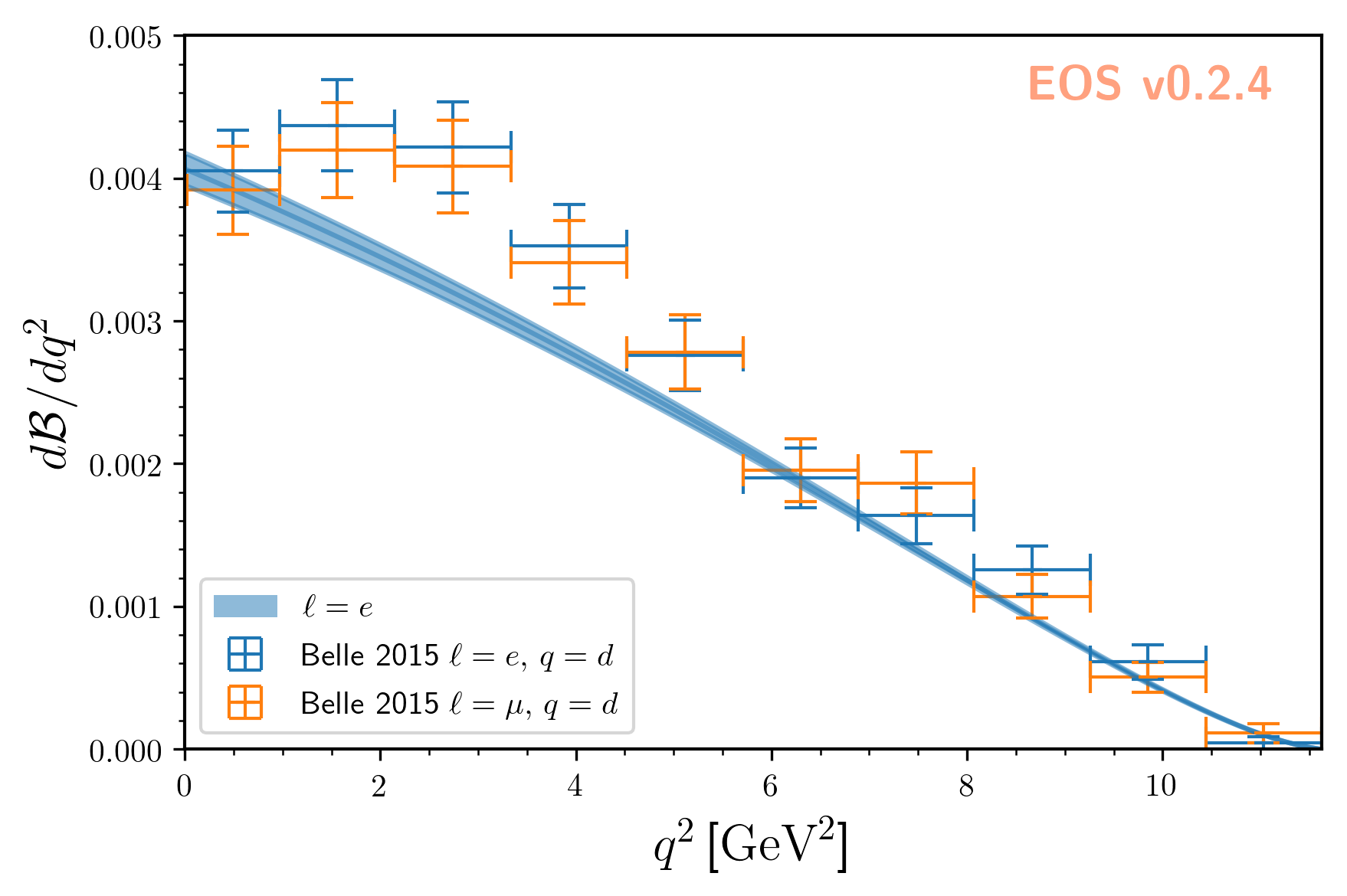}
\caption{Two of the plots created by the example notebooks.
Left: Histogram of random variates for $|V_{cb}|$ obtained and plotted with the \EOS Python interface, overlaid with a kernel density estimate of the posterior probability density.
Right: Uncertainty envelope at $68\%$ credibility of the posterior-prediction for the $\bar{B}\to De^-\bar\nu$ differential branching fraction as a function
of $q^2$, the $e^-\bar\nu$ mass squared. The curves and bands are overlaid with data points by the Belle collaboration from measurements of the
$\bar{B}\to D e^-\bar\nu$ and $\bar{B}\to D\mu^-\bar\nu$ branching fraction in $q^2$ bins.
}
\label{fig:inference}
\end{figure}

The Python3 interface is recommended to all novice users, and example notebooks using the Jupyter software \cite{Jupyter}
are available \cite{EOS-examples}. These examples showcase how to use \EOS to carry out typical tasks to work on any of the use cases
listed above.
The examples use semileptonic decays of the type $\bar{B}\to D\ell\bar\nu$ to illustrate how to predict
integrated and differential kinematical distributions, infer the SM parameter $|V_{cb}|$ and hadronic parameters,
and generate pseudo events within the SM and for BSM benchmark points.
A detailed write-up with additional explanations can be viewed on the \EOS webpage page \cite{EOS-doc}.
In figure~\ref{fig:inference}, two example plots are presented that illustrate the capabilities of \EOS
with the Python3 interface.

The \EOS developers invest effort on providing precise and accurate
predictions of flavour observables in and beyond the SM. Flavour observables
are implemented independent of any concrete BSM model by using process-specific
Weak Effective Theories (WETs). In this approach, a variety of hadronic matrix of local WET
operator elements are needed. \EOS allows to switch between different approaches for the
evaluation of these hadronic matrix elements at run time.
To connect low-energy flavour observables within their various WETs
to one common high-energy BSM scenario within the Standard Model Effective Field Theory (SMEFT), matching and running
between both theories is needed. To this end, \EOS interfaces with the Wilson Python package \cite{Aebischer:2018bkb}
via the Wilson Coefficients exchange format \cite{Aebischer:2017ugx}.



\newpage
\talk{\texttt{flavio} and \texttt{smelli}}{David M. Straub}{Excellence Cluster Universe/TUM}

\noindent
\texttt{flavio} \cite{Straub:2018kue,flavioweb} and \texttt{smelli} \cite{Aebischer:2018iyb,smelliweb}
are two closely related open source Python packages.

\texttt{flavio} has the following main features:
\begin{itemize}
\item It is a general observable calculator (with uncertainties) in terms of WET or SMEFT Wilson coefficients,
\item it contains a database of experimental meausurements,
\item it allows the automated construction of likelihoods.
\end{itemize}
In addition, it contains convenient plotting routines, interfaces to fitters (MCMC), and a frequentist likelihood profiler.

\texttt{smelli}, the
\textbf{SME}FT \textbf{L}ike\textbf{li}hood package, is built on top of \texttt{flavio} and provides a global likelihood
in the space of SMEFT Wilson coefficients. The main motivation for \texttt{smelli} is:
\begin{itemize}
\item providing a consistent set of observables included in the likelihood,
\item correct treatment of SM parameters in the presence of $D=6$ effects,
\item construction of a nuisance-free likelihood,
\item more informative presentation of results (table of observables with pulls etc.).
\end{itemize}

While \texttt{flavio} aims to be easy to modify and very flexible,
the focus in \texttt{smelli} is on the ease of use and consistency rather than
generality.

Thanks to the wilson package (Section~\ref{sec:wilson}), \texttt{flavio}, which originally
started as a pure flavour physics package, now also supports electroweak
precision tests and will soon add Higgs physics. In principle, every
observable where new physics enters via Wilson coefficients of local
operators can be added. The long term goal is to include all processes
sensitive to dimension-6 SMEFT operators.
Correspondingly, the long-term goal for \texttt{smelli} is to become truly \textit{global} in constraining as many directions in SMEFT Wilson coefficient space as at all possible.

As of version 1.5, \texttt{flavio} includes the following classes of observables:
\begin{itemize}
\item $B$ physics: $B\to (V, P, X)\ell\ell$, $B\to \ell\ell$, $B\to (V, X)\gamma$, $\Lambda_b\to\Lambda\ell\ell$, $B\to (V, P, X)\ell\nu$, $B\to \ell\nu$, mixing
\item $K$ physics: $K\to\pi\nu\nu$, $K\to \ell\ell$, $K\to \ell\nu$, $K\to \pi\ell\nu$, $\epsilon_K$, $\epsilon'/\epsilon$
\item $D$ physics: $D\to \ell\nu$, CPV in mixing
\item $\mu$ physics: $\mu\to e\gamma$, $\mu\to 3e$, $\mu$-$e$ conversion, $\nu$ trident
\item $\tau$ physics: $\tau\to 3\ell$, $\tau\to \ell\gamma$, $\tau\to(P,V)\ell$, $\tau\to V\nu$, $\tau\to\ell\nu\nu$
\item EWPT: All LEP-1 $Z$ and $W$ pole observables
\item Dipole moments: $(g-2)_{e,\mu,\tau}$, $d_n$
\end{itemize}
Near-future versions will add nuclear and neutron $\beta$ decays
as well as Higgs production and decay.

\begin{figure}
  \includegraphics[width=\textwidth]{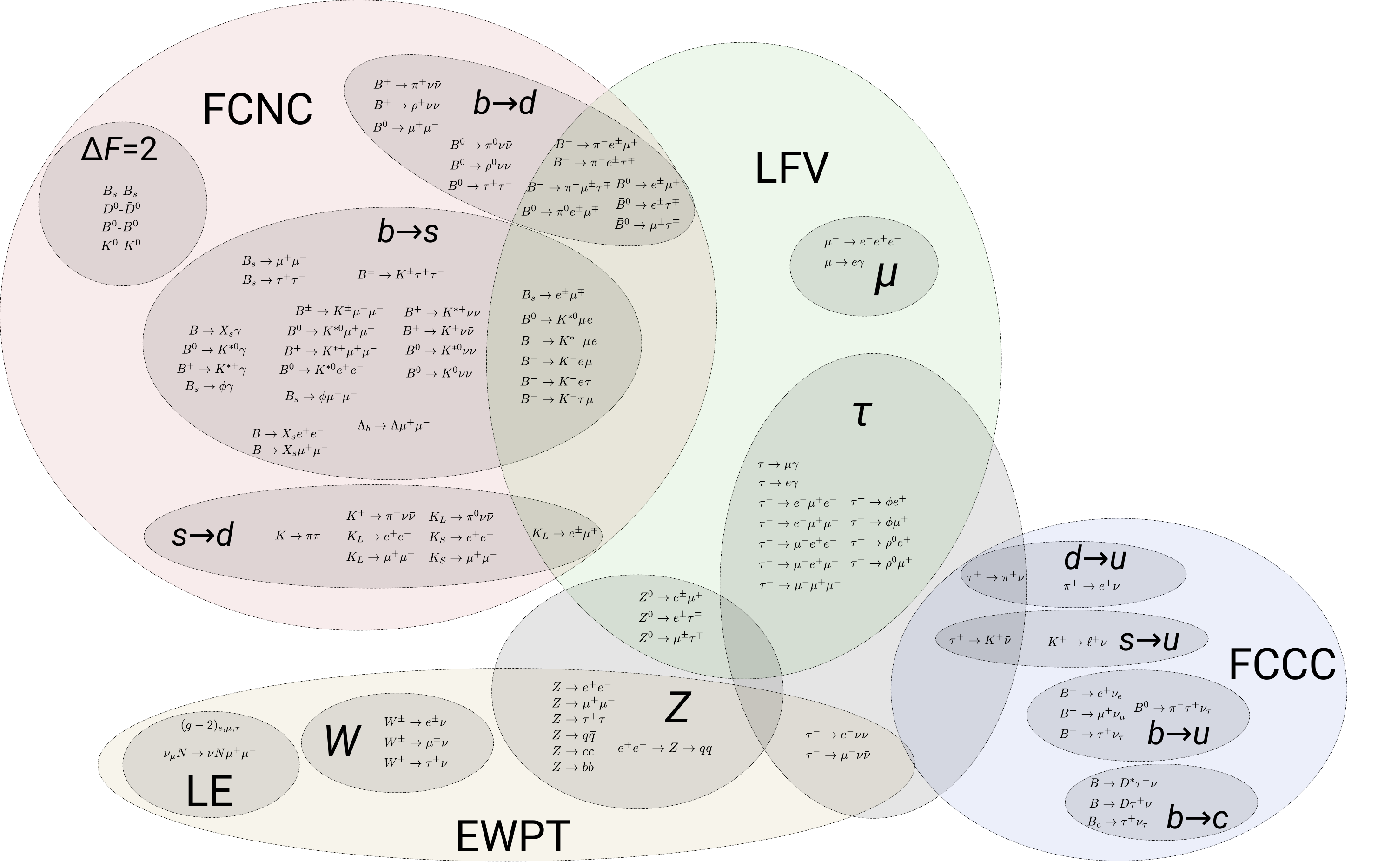}
  \caption{Sketch of observables currently included in \texttt{smelli}'s
  global SMEFT likelihood.}
  \label{fig:smelli-obs}
\end{figure}

\texttt{smelli}, as of version 1.3, includes every observable in \texttt{flavio} that fulfills two criteria: it has been measured and it is relevant for constraining SMEFT Wilson coefficients.
The classes of observables are visualized in the sketch in figure~\ref{fig:smelli-obs}.
There are only two limitations at present. One is that
semi-leptonic charged-current meson decays
are not included yet, as the effect of new physics on CKM element
extractions is not treated consistently yet.
However, a solution similar to
\cite{Descotes-Genon:2018foz}
has been implement in \texttt{smelli} and will be public soon.
The second limitation is that the statistical approach used to be
able to provide a \textit{nuisance-free} likelihood
(for details see \cite{Aebischer:2018iyb})
does not allow to include observables where theory uncertainties are
strongly dependent on new physics, as is the case e.g.\ for the
neutron EDM or CP violation in $B\to DK$ decays.
Ideas to solve this second limitation are being explored.

\begin{figure}
  \centering
  \includegraphics[width=0.65\textwidth]{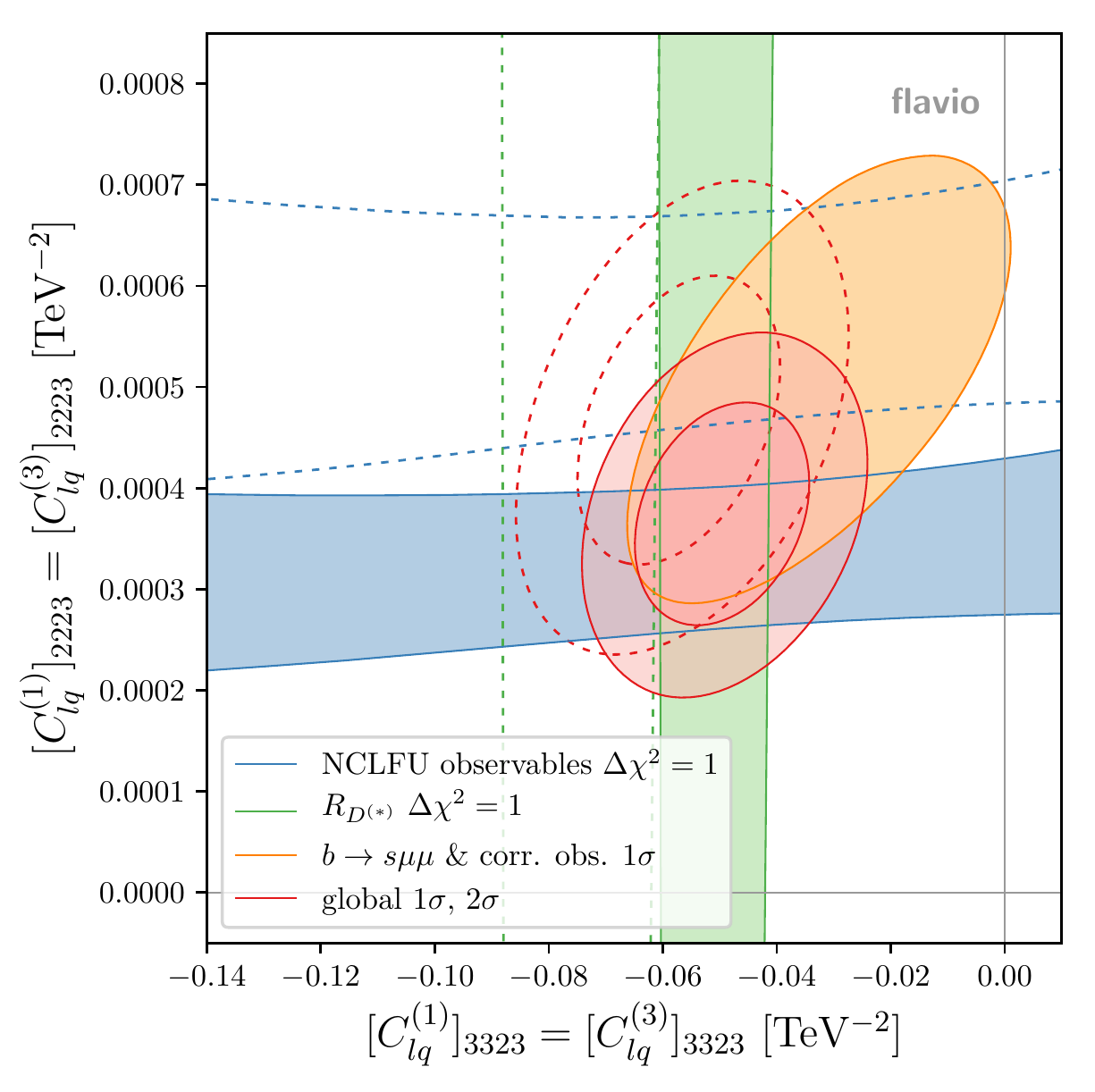}
  \caption{\texttt{smelli} global likelihood (and sub-likelihoods)
  in the plane of two SMEFT Wilson coefficients at the scale 2~TeV,
  plotted with \texttt{flavio.plots}. Taken from \cite{Aebischer:2019mlg}.
  }
  \label{fig:smelli-plot}
\end{figure}

A brief interactive tutorial demonstrating the usage and main features of
\texttt{flavio} and \texttt{smelli} can be found at the following URL:

\url{https://github.com/DavidMStraub/flavio-smelli-mini-tutorials}

An example of a possible application of the two codes is given
by the well-known plot in figure~\ref{fig:smelli-plot},
showing the interplay of charged- and neutral-current $B$ decays
on semi-leptonic SMEFT Wilson coefficients,
very relevant for the present $B$ ``anomalies'' \cite{Aebischer:2019mlg}.

\newpage
\talk{Matching the flavour symmetric SMEFT to flavour observables}{Sophie Renner}{SISSA International School for Advanced Studies}

\noindent
This talk was based on Ref.~\cite{Hurth:2019ula}.

\subsection{Introduction}


The success of the CKM picture of quark flavour mixing points towards there being no large sources of flavour breaking beyond the Standard Model (SM). If TeV-scale new physics exists, models with a flavour symmetry are therefore highly motivated. In the following I describe calculations of some of the most important flavour effects generated by flavour symmetric operators within the Standard Model Effective Field Theory (SMEFT).

%

\subsection{\boldmath{Matching the $U(3)^5$ invariant SMEFT to WET coefficients up to one-loop}}
We select operators in the Warsaw Basis~\cite{Grzadkowski:2010es} of dimension 6 operators by starting from a $U(3)^5$ flavour symmetry defined as
\begin{equation}
U(3)_q \times U(3)_{u} \times U(3)_{d} \times U(3)_l \times U(3)_{e},
\end{equation}
which is the largest global symmetry of the gauge sector of the SM Lagrangian, and under which the SM fermion fields transform in the fundamental of their corresponding $U(3)$, and as singlets under the others.
We consider only the effects of operators which are overall singlets under this $U(3)^5$ symmetry, an assumption which essentially eliminates tree level flavour changing neutral currents (FCNCs). The motivations for this are as follows
\begin{itemize}
\item Will allow incorporation of flavour data into global SMEFT fits, which often use the same flavour assumption
\item Approximates a ``worst-case scenario'' for the effects of TeV-scale new physics in flavour measurements, and thus represents an estimate of the irreducible flavour effects that might be expected 
\item Provides a starting point from which to explore other motivated flavour symmetries (e.g.~less restrictive MFV scenarios, $U(2)^5$ in the first two generations, etc)
\end{itemize}

\begin{figure}
\begin{center}
\includegraphics[width=4cm]{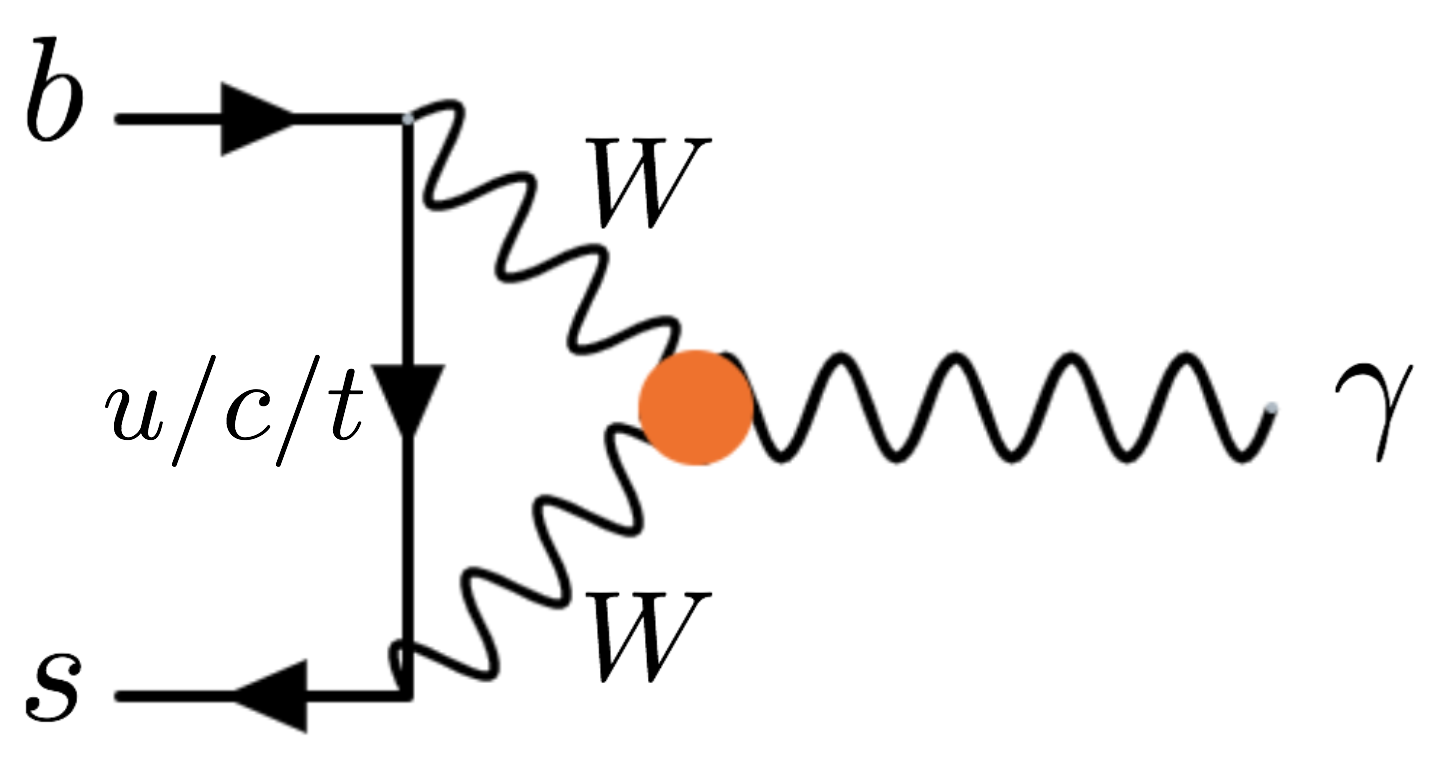}~~
\includegraphics[width=4.5cm]{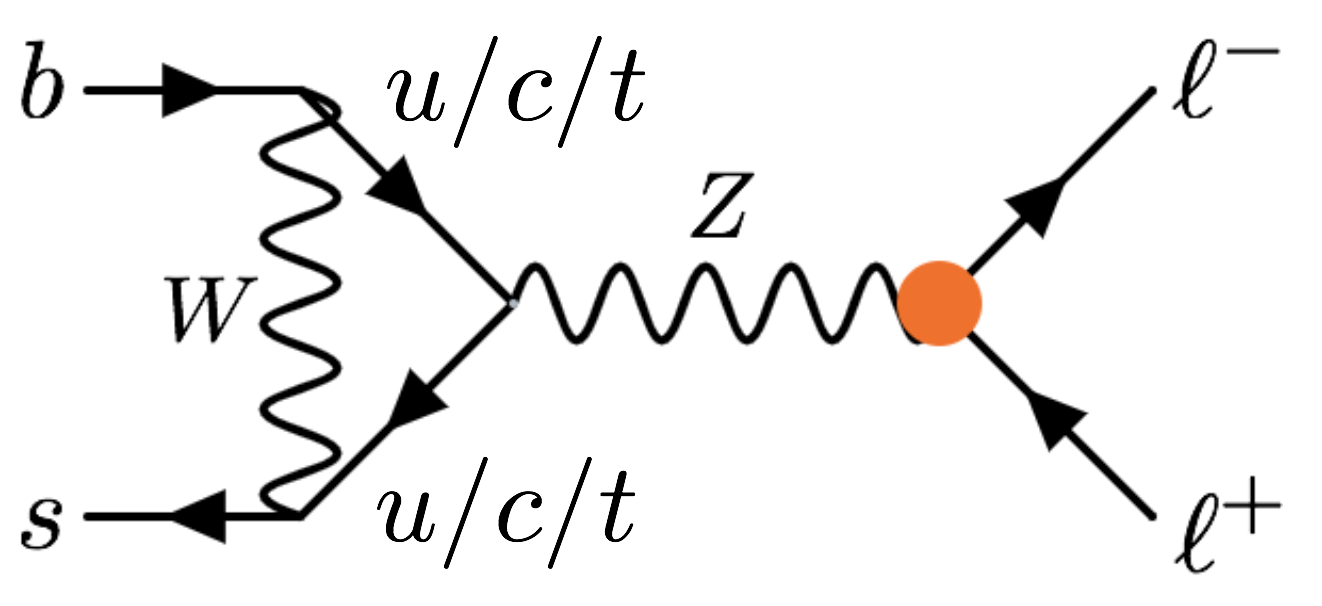}~~
\includegraphics[width=3.5cm]{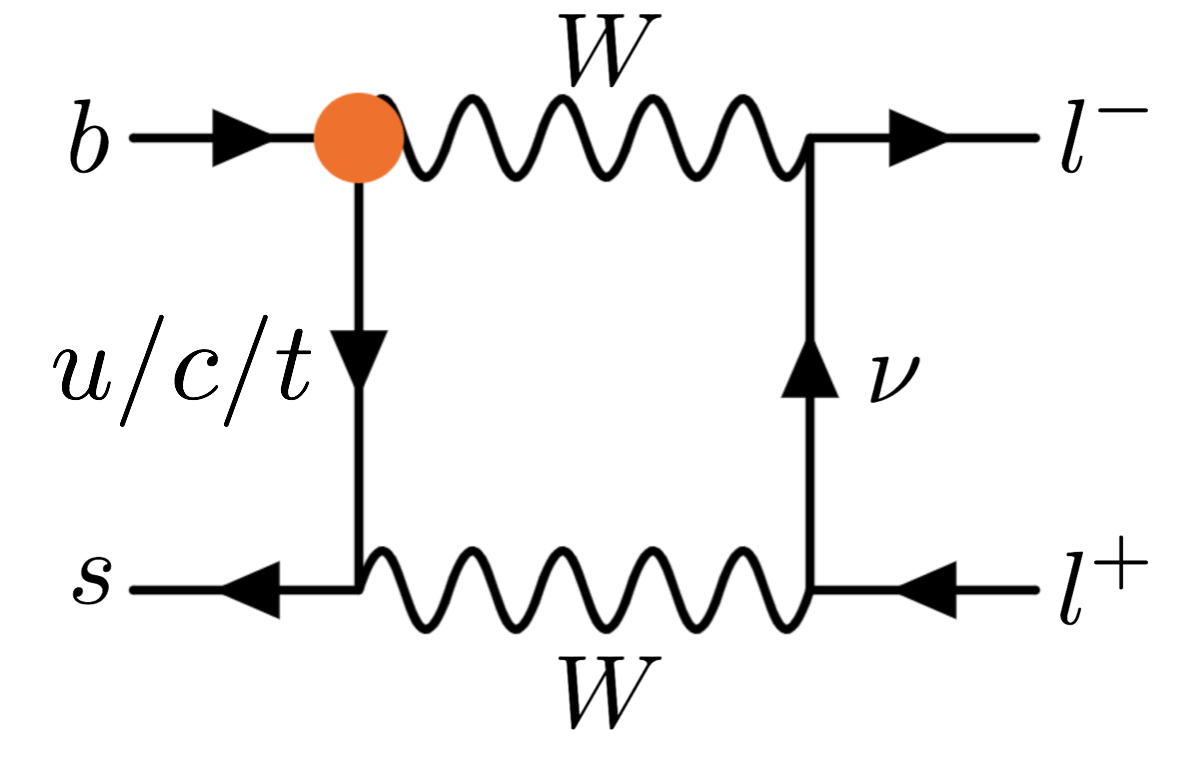}
\caption{Examples of diagrams for one-loop matching, where the orange blobs represent insertions of SMEFT operators}
\label{fig:diagrams}
\end{center}
\end{figure}

We match onto operators in the Weak Effective theory (WET) which generate some of the most sensitive flavour observables to new physics: semileptonic down-type FCNC decays (e.g.~$B\to K^{(*)} l^+ l^-$) and down-type meson mixing (e.g~$B_s$-$\bar{B}_s$, $K^0$-$\bar{K}$). 
The flavour symmetry assumption ensures that there are no tree-level FCNCs induced by the SMEFT operators we consider,\footnote{arguably with the exception of the  $Q_{qq}^{(1,3)}$  operators; see \cite{Hurth:2019ula} for a discussion of these} so matching at one-loop is necessary, with the flavour change arising from SM loops involving $W^{\pm}$ bosons. Examples of the diagrams involved are shown in Figure~\ref{fig:diagrams}.

\subsection{Results and discussion}
For results we refer to \cite{Hurth:2019ula}. Here we point out a few general properties of the calculations. The $U(3)^5$ flavour symmetry we assume ensures that the results of the matching calculations share many properties with the SM, in particular:
\begin{itemize}
\item \emph{GIM mechanism}: all results depend on $m_t$
\item \emph{No right-handed currents}: the matching only affects the coefficients of WET operators containing left-handed light quarks
\item \emph{Same CKM factors}: the matching produces effects with the same CKM suppressions as in the SM
\end{itemize}
Some of the operators we consider also enter into observables that are measured to fix the input parameters of the theory~(see e.g.~\cite{Han:2004az,Falkowski:2014tna,Berthier:2015oma,Brivio:2017bnu}). We take these effects into account such that the results are written in terms of measured parameters, and present our results in two common schemes for the inputs fixing the electroweak sector of the theory: $\lbrace m_W,m_Z,G_F\rbrace$ and $\lbrace \alpha,m_Z,G_F\rbrace$. Where possible, we compared results to those obtained previously in the literature \cite{Aebischer:2015fzz,Drobnak:2011wj,Drobnak:2011aa,Bobeth:2015zqa,Endo:2018gdn,Grzadkowski:2008mf}. These calculations open up new possibilities for constraining the coefficients of SMEFT operators with flavour data.



\newpage
\talk{MatchMaker}{Jos\'e Santiago\footnote{In collaboration with C. Anastasiou, A. Carmona, A. Lazopoulos}}{CAFPE and University of Granada}

\subsection{Introduction}

One of the (many) advantages of effective field theories (EFT) is that the
comparison between experimental data and their implication in new
physics models can be split in two independent steps. The bottom-up
approach allows to parametrize in a maximally agnostic way the
experimental data in terms of the Wilson coefficients of the
corresponding EFT, without any relation to specific new physics
models. The result is a global likelihood that can be computed thanks to the
tools described elsewhere in this document. The top-down approach on
the other hand introduces the necessary model discrimination by
computing the Wilson coefficients in terms of the parameters of the
new physics model.

The beauty of EFT is that it provides a power counting rationale to
organize the model dependence inherent to the top-down approach, as
the number of classes of models that contribute at a certain order in
the perturbative expansion (loops and operator dimension) is finite.
In this spirit, the complete classification of arbitrary extensions of
the Standard Model (SM) that include new particles of spin smaller than 3/2
and contribute at tree level at the SM effective field
theory (SMEFT) of dimension 6 has been recently
achieved~\cite{deBlas:2017xtg}, building on previous partial
efforts~\cite{deBlas:2014mba,delAguila:2010mx,delAguila:2008pw,delAguila:2000rc}. Furthermore,
technology allows us to automate this top-down
approach~\footnote{As an example
  \texttt{MatchingTools}~\cite{Criado:2017khh} was extensively used
  in checking the results of~\cite{deBlas:2017xtg}.}, a necessary
ingredient if we want to extend this complete classification beyond
the tree-level approximation. This motivated us to develop \mm, an
automated tool to perform tree-level and one loop matching of
arbitrary new physics models to the SMEFT.

\subsection{\mm: general philosophy}

New physics models can be matched to an EFT either via functional
methods, literally integrating out the heavy degrees of freedom
directly in the path integral
(see~\cite{Henning:2014wua,Drozd:2015rsp,delAguila:2016zcb,
  Henning:2016lyp,Fuentes-Martin:2016uol,Zhang:2016pja,Ellis:2017jns}
 for recent progress in this direction) or
via a diagrammatic approach. This latter approach, in which off-shell
1-light-particle-irreducible (1lPI) Green
functions are compared in the full model and the EFT,  is the one used
in \mm.

\mm relies on well established methods and tools for the process of
tree-level and one-loop matching. It consists of a \python engine,
which ensures that it is cross-platform, easy to install and
flexible, and the following standard tools for the different steps of
the calculation:
\begin{itemize}
  \item \feynrules~\cite{Alloul:2013bka} is used to define the new
    physics model and to
    automatically compute the corresponding Feynman rules. Flavor
    indices are implicit dummy variables all through the calculation
    and therefore the number of generations is arbitrary with no
    computation penalty.
  \item \qgraf~\cite{Nogueira:1991ex} is used to generate all the
    relevant amplitudes in the full and effective theories. These
    amplitudes are automatically dressed by \mm with the Feynman rules
    computed in the previous step.
    \item The actual calculation of the corresponding amplitudes is
      performed with \form~\cite{Kuipers:2012rf}. This includes
      external momentum expansion, tensor reduction, partial
      fractioning, Dirac algebra and integration by part
      identities. All the calculations are done in dimensional
      regularization following the $\overline{\mbox{MS}}$ renormalization
      scheme.
      \item Finally, the comparison of the amplitudes in the full and
       effective theories is performed using \mathematica and the
       final form of the Wilson coefficients is stored in a file.
\end{itemize}

A few comments regarding the procedure described above are in
order. As we have mentioned, we perform an off-shell matching in which
the external particles are not required to be on-shell (only full
momentum conservation is imposed). The rationale behind this choice is
that we can restrict the calculation to 1lPI Green functions, as
opposed to full S-matrix elements, and that the full off-shell
kinematic structure provides a highly non-trivial kinematic redundancy
that we use to cross-check the results obtained by \mm. The down-side
of this choice is that redundant operators, those that can be
eliminated by field redefinitions and do not contribute to physical
observables, have to be included in the process of matching.
In order to keep the number of operators to consider under control we
use the background field version of the 't Hooft-Feynman
gauge~\cite{Abbott:1980hw}  so that
only gauge invariant operators have to be considered. Following our
philosophy of maximum flexibility we have not fixed a particular
basis of evanescent operators, leaving the user to choose such a
basis. The results of the matching are given
first in the full off-shell Green basis, including all the relevant
redundant and evanescent operators. This also includes the matching
effects on the effective operators of dimension smaller than 6
(including the operators in the SM Lagrangian). On top of this
completely general result for the tree-level and one-loop matching,
\mm also provides the result of the matching in the Warsaw
basis~\cite{Grzadkowski:2010es}, assuming 4-dimensional properties for
the gamma matrices to reduce the evanescent operators. The elimination of the
redundant operators to the Warsaw basis is performed by means of the
equations of motion of the dimension-4 Lagrangian, after canonical
normalization of the fields and the inclusion of the corrections to
this Lagrangian from the matching process.

\mm uses the following tests to check the correctness of the produced
results:
\begin{itemize}
\item Full off-shell kinematic and gauge dependence. Off-shell
  kinematics provides a powerful and highly non-trivial test of the
  matching. Similarly, all components in gauge space are independently
  checked.
  \item Ward identities. We carefully compare amplitudes with
    different number of external legs, in which a momentum is replaced
    with the corresponding gauge bosons, to check all the relevant
    Ward identities.
    \item Symmetry and hermiticity properties of the Wilson
      coefficients. Some Wilson coefficients have symmetry
      properties under the exchange of flavor indices, including in
      cases complex conjugation. These properties are systematically
      checked in the result of the matching.
\end{itemize}

\subsection{Status and future prospects}

At the time of this writing \mm is not yet publicly available. We are
finalizing the last checks of the program and we expect to make it
public in the near future. The current version produces in an
automated way the tree-level and one-loop matching of an arbitrary new physics
model (with the only restriction that it has to be implementable in
\feynrules) into the SMEFT, provided no fermion-number violating
couplings are introduced in the model. We expect the latter to be
handled in future versions of the program. The grand goal is to extend
this tool to arbitrary effective Lagrangians.



\newpage
\talk{Lepton dipole moments at two loops in the SMEFT
}{Giovanni Marco Pruna}{INFN, Sezione di Roma Tre
}

\noindent
In the last decade, the absence of signals for physics beyond the Standard Model (BSM) at collider experiments, together with the increasing precision of both high- and low-energy experiments, corroborated the idea that there can be a considerable scale separation between the SM and New Physics, thus creating strong grounds for deeper studies of the Standard Model Effective Field Theory (SMEFT).

In fact, even though the SMEFT was introduced several decades ago~\cite{Buchmuller:1985jz}, a systematic treatment of such theory above and below the electroweak symmetry-breaking (EWSB) scale was completed only a few years ago~\cite{Grzadkowski:2010es,Lehman:2014jma,Jenkins:2017jig}.

A similar fate occurred to the study of quantum fluctuations in SMEFT. Although many partial results had been presented in literature, a methodical analysis of the one-loop anomalous dimensions was only recently performed~\cite{Jenkins:2013zja,Jenkins:2013wua,Alonso:2013hga,Liao:2016hru,Jenkins:2017dyc,Liao:2019tep}.

Beyond the one-loop level, there are not many results, and these are not organised in a structured catalogue~\cite{Altarelli:1980fi,Buras:1989xd,Buras:1991jm,Ciuchini:1992tj,Ciuchini:1993vr,Misiak:1994zw,Chetyrkin:1996vx,Chetyrkin:1997fm,Ciuchini:1998ix,Buras:2000if,Gracey:2000am,Gambino:2003zm,Gorbahn:2004my,Gorbahn:2005sa,Degrassi:2005zd,Czakon:2006ss}. However, the precision level that will be reached by future experiment will require a consistent knowledge of the two-loop leading contributions in SMEFT. Therefore, a collaborative effort is required to reach this goal in a reasonable time. It is beyond the scope of this document to analyse the phenomenological implications of a thorough knowledge of the two-loop anomalous dimensions in SMEFT. The focus instead is on the characteristic case of lepton dipole moments.

In the near future, a worldwide experimental plan will test such observables with unprecedented sensitivity. In two-to-five years, lepton-flavour violation (LFV) will be investigated in the muon sector with an increase of three orders of magnitude in sensitivity at the MEG~II (PSI)~\cite{Baldini:2013ke}, Mu3e (PSI)~\cite{Blondel:2013ia}, COMET (J-PARC)~\cite{Cui:2009zz} and Mu2e (FNAL)~\cite{Carey:2008zz,Kutschke:2011ux} experiments. ACME~II at Harvard University delivered a new result~\cite{Andreev:2018ayy} on the electric dipole moment (EDM) of electrons with a significantly improved sensitivity. This year, the Muon $g-2$ experiment (FNAL)~\cite{Grange:2015fou,Gohn:2017dsp} will also deliver exciting results on the anomalous magnetic moment of the muon ($a_\mu$), possibly addressing the nature of the long-standing discrepancy between measured and predicted values.

In this context, it has already been shown that a proper evaluation of the quantum fluctuations in SMEFT gives a deeper understanding of the phenomenological implications in searches for new physics~\cite{Crivellin:2013hpa,Pruna:2014asa,Pruna:2015jhf,Davidson:2016utf,Davidson:2016edt,Cirigliano:2017azj,Crivellin:2017rmk,Davidson:2017nrp,Pruna:2017tif,Crivellin:2018ahj,Davidson:2018rqt,Davidson:2018kud,Dekens:2018pbu,Panico:2018hal}.

However, apart from the fact that evaluations of further loop levels would guarantee more control over the phenomenological interpretation, there is a formal subtlety that calls for a systematic determination of the two-loop leading contribution to the evolution/mixing of SMEFT operators.

In fact, it is well-known that matching an ultraviolet (UV) complete theory with its low-energy effective representation can distribute the leading contributions in different perturbative orders: the pioneering work on $b\rightarrow s\gamma $ transitions performed in the $1990$s revealed that this is a standard feature of the fermion dipole operators when quantum fluctuations are evaluated in an effective field theory~\cite{Ciuchini:1993ks,Misiak:1993es}.

This aspect implies that the matching procedure performed at the one-loop level does not provide meaningful phenomenological information unless accompanied by a consistent evaluation of the anomalous dimensions at the two-loop level. This issue is made more radical by the fact that both the matching coefficient and the anomalous dimensions can depend on the regularisation scheme used to perform the aforementioned computation.

Therefore, the only way to remove all the scheme ambiguities from the correct phenomenological interpretation is to compute the anomalous dimensions at the two-loop level and the matching coefficients at the one-loop level in SMEFT, potentially in different regularisation schemes to cross-check the final outcome.

Other ambiguities can arise with the treatment of the antisymmetric Levi-Civita tensor in dimensional regularisation, and a careful implementation of the evanescent operator technology~\cite{Altarelli:1980fi,Buras:1989xd,Ciuchini:1993vr} is required.

The current level of automation allows implementation of the SMEFT Lagrangian in tools that can generate the relevant set of Feynman rules (\emph{e.g.} {\tt FeynRules~v2.3}~\cite{Alloul:2013bka}) to be interpreted by packages for multi-loop computations (\emph{e.g.} {\tt FeynArts~v3.11}~\cite{Hahn:2000kx} and {\tt FormCalc~v9.8}~\cite{Hahn:1998yk,Hahn:2016ebn}). However, a completely automated chain is not yet available. For example, following the package chain that was just presented, one should notice that FormCalc struggles to deal with four-fermion operators. Therefore an intermediate Form file should be retained to be further elaborated off line in {\tt Form~v4.2}~\cite{Kuipers:2012rf}.

Although it is not perfect, this strategy allows the non-integrated amplitudes to be further related to any effective operator by a projection algorithm. Dipole operators can be extracted with an off-shell projection~\cite{Ciuchini:1997xe}.

After this, extracting the UV poles will require a strategy to regularise potential infrared (IR) divergences. In recent years, many techniques have been developed to this. However, all of these techniques involve rearranging the propagators into a tadpole integral, followed by a truncation that takes into account the superficial degree of divergence of the loop integral. One convenient choice is to rearrange every propagator with an IR mass regulator exactly~\cite{Misiak:1994zw,Chetyrkin:1997fm,Cherchiglia:2010yd}. Even though this approach is very efficient, it has the disadvantage that to be consistent one must add the new regulator mass in the original Lagrangian (possibly interfering with pre-existing masses) and carry out the potential set of counterterms that arise because of this insertion. Another choice is to add an artificial mass term to the massless gauge bosons and then expand the integral in the external momenta~\cite{Ciuchini:1997xe} (which is equivalent to an IR rearrangement into a multi-massive tadpole). Since two-loop tadpoles are very well-known objects~\cite{Davydychev:1992mt,Bogner:2017xhp}, this second method will also give a linear outcome. Hence the two methods can be considered equivalent.

The approach described in this section successfully reproduces the two-loop anomalous dimensions in $b\to s\gamma$ transitions below the EWSB scale in conventional dimensional regularisation~\cite{Ciuchini:1993fk}, and can be adapted to compute the two-loop anomalous dimensions of lepton dipole moments with the operator basis presented in~\cite{Jenkins:2017jig}. The result will appear in~\cite{Pruna:2018xmd}.

At current the state of the art, the procedure described here is surely one of the most efficient. However, there is no guarantee that it cannot be surpassed by interesting alternatives~\cite{deVries:2019nsu}.



\newpage
\talk{NLO corrections to $h\rightarrow b\bar{b}$ in the SMEFT}{Darren Scott}{University of Amsterdam/Nikhef}

\noindent
Performing higher order perturbative corrections within the SMEFT can be important for at least two reasons. Not only do such corrections reduce the residual scale dependence in the resulting predictions as in the SM, but new Wilson coefficients can appear at NLO which are not present in the tree level result. On occasion these Wilson coefficients can have large numerical prefactors making them an important contribution when incorporated in a global fit. Some of these large contributions would not necessarily be captured by a standard RGE analysis thus making it necessary to compute the NLO corrections directly in order to properly capture the impact of such coefficients. This talk introduces NLO corrections to the decay $h\rightarrow b\bar{b}$ in the dimension-6 SMEFT~\cite{Gauld:2015lmb,Gauld:2016kuu,Cullen:2019nnr}, with emphasis on a number of complications or subtleties in the calculation of the dimension-6 decay rate which are not seen in the SM. We focus in particular on the renormalization of the electric charge, large NLO corrections (and the choice of renormalization scheme), and the use of decoupling relations. Further details can be found in~\cite{Cullen:2019nnr} as well as the complete answer for the decay rate in \texttt{Mathematica} notebooks available with the arXiv submission.

In the on-shell scheme the electric charge is defined to be exactly equal to the tree level coupling of the $\gamma ff$-vertex at zero momentum transfer to the photon. The electric charge is universal, so the choice of $f$ does not affect the result one gets for the form of the counterterm $\delta e$. It can be shown that $\delta e$, using electroweak SM Ward identities, can be written (see e.g.~\cite{Denner:1991kt}\footnote{The difference in sign on the second term comes from a different sign convention used in \cite{Denner:1991kt} for the covariant derivative.})
\begin{equation}
\label{eq:eRen_SM}
\frac{\delta e}{e} = \frac{1}{2}\frac{\partial \Sigma_T^{AA}(k^2)}{\partial k^2}\bigg|_{k^2=0} - \frac{(v_f-a_f)}{Q_f}\frac{\Sigma_T^{AZ}(0)}{M_Z^2} \, ,
\end{equation}
where $\Sigma_T^{AB}$ is the transverse component of the $A\rightarrow B$ 2-point function, $v_f$ and $a_f$ are the vector and axial couplings of fermion $f$ to the Z-boson, and $Q_f$ is the electric charge of the fermion in units of $e$. In the SM, $v_f-a_f = -2Q_f \hat{s}_w/\hat{c}_w$ (where $\hat{s}_w$ and $\hat{c}_w$ are the sine and cosine of the weak mixing angle) making it clear that eq.~(\ref{eq:eRen_SM}) becomes independent of the choice of fermion used to compute $\delta e$.
The na\"ive extension of eq.~(\ref{eq:eRen_SM}) to dimension-6 (i.e. including dimension-6 effects in $\Sigma_T^{AB}$, $v_f$, and $a_f$) leads to seemingly contradictory results. Specifically, for operators of the form $C_{Hf}(H^\dagger i \overleftrightarrow{D}_\mu H)(\bar{f}\gamma^\mu P_R f)$ where $H$ is the Higgs doublet with vacuum expectation value $\hat{v}_T$, one finds $v_f^{(6)}=-a_f^{(6)}=C_{Hf}\hat{v}_T^2/4\hat{c}_w\hat{s}_w$. Thus, this na\"ive extension leads to the result that the electric charge renormalization depends on fermion charge $Q_f$.\footnote{One can show that a SM Ward identity used to derive eq.~(\ref{eq:eRen_SM}) is actually not satisfied by such operators, and so presumably receives corrections at dimension-6.} One can of course avoid this issue by renormalizing
the $\gamma f f$-vertex directly, without using the SM Ward identities, but it would be interesting to derive an all-orders equation analogous to eq.~(\ref{eq:eRen_SM}) which correctly takes into account dimension-6 terms.

Another issue is related to the origin of large NLO corrections in the results and their relation to the choice of renormalization scheme. We identify two such contributions; firstly those from QCD/QED type diagrams (those containing at least one gluon/photon) and secondly those from tadpole graphs containing a top quark or, to a lesser extent, a heavy boson. The former are smaller if one uses the $\overline{\text{MS}}$ scheme for the $b$-quark mass\footnote{In fact even in this scheme there is still a large contribution to the $C_{HG}$ coefficient (operator: $HHGG$) from a double logarithm in $m_b/m_H$. Using the $\overline{\text{MS}}$ scheme still reduces the size of the corrections and removes the single logarithmic contribution as in the SM however.}, while on the other hand tadpole contributions can be made to vanish if one instead uses the on-shell scheme everywhere. A solution to this apparent dilemma is through the use of decoupling relations. That is, we work in a scheme where the $b$-quark mass (and electric charge $e$) are defined in a low energy QCD$\times$QED theory with the top quark and heavy bosons decoupled (see e.g.~\cite{Bednyakov:2016onn}). We write the relation between the full theory $b$-quark mass in the $\overline{\text{MS}}$ scheme $\overline{m}_b(\mu)$ and the mass as defined in the low energy QCD$\times$QED theory $\overline{m}^{(\ell)}_b(\mu)$ as
\begin{equation}
\overline{m}_b(\mu) = \zeta_b(\mu,m_t,m_H,M_W,M_Z)\overline{m}^{(\ell)}_b(\mu) \, ,
\end{equation}
where the decoupling constant $\zeta_b$ captures the contributions from the top quark and heavy bosons to the $b$-quark mass. The decoupling constants can be calculated perturbatively through one-loop matching. The net effect is that the contributions from top quarks and heavy bosons are calculated in the on-shell scheme where the tadpole graphs cancel, while the QCD/QED contributions are calculated in the $\overline{\text{MS}}$ scheme where their contributions are smaller.

In this short text we have covered some issues related to electric charge renormalization and the use of decoupling relations when using a mix of on-shell and $\overline{\text{MS}}$ renormalization schemes to avoid anomalously large NLO corrections. There are some other subtleties not touched on here, such as Higgs-Z mixing, scale uncertainties, the flavour structure (see also~\cite{Brivio:2017btx,Alonso:2013hga}), and gauge fixing (see also~\cite{Dedes:2017zog,Helset:2018fgq,Misiak:2018gvl}) but which can also be found in the full paper~\cite{Cullen:2019nnr} along with additional details on the points raised here.



\newpage
\talk{(SM)EFT, thoughts about what everybody has seen}{Giampiero Passarino}{
Dipartimento di Fisica Teorica, Universit\`a di Torino\\
INFN, Sezione di Torino}

\noindent
There is increasing need to assess the impact and the interpretation of $\mrdim = 6$ and $\mrdim = 8$ operators within
the context of the Standard Model Effective Field Theory (SMEFT~\cite{Brivio:2017vri}). The observational and
mathematical consistency of a construct based on $\mrdim = 6$ and $\mrdim = 8$ operators should be critically
examined~\footnote{There are no eternal facts, as there are no absolute truths.}
in the light of known theoretical results~\cite{Passarino:2019yjx}, including: local, non{-}local, hard and soft terms
or why we should not forget loops~\cite{Passarino:2016pzb} and Landau singularities; mixing, or why SMEFT may not be
as general as we think; linear vs. quadratic EFT representations.

\subsubsection*{Non{-}local  and all that} 

Consider the following scenario: the Standard Model (SM), valid for $\mrE \ll \Lambda$, the corresponding
EFT extension (say SMEFT) and $X$, an UV completion of the SM (or the next theory in a tower of effective theories);
we are interested in matching the low $\mrE$ limit of $X$ to the (SM)EFT.
Non{-}local effects correspond to long distance propagation and hence to reliable predictions at low energy,
local terms by contrast summarize the unknown effects from high energy: having both local and non{-}local terms allows
us to implement the full (one-loop) EFT program~\cite{Donoghue:2017pgk}. Heavy{-}light terms describe a multi scale
scenario: the light masses ($m_i$), the Mandelstam invariants characterizing the process ($s_{i j \dots k}$) and the
heavy scale ($\Lambda$),
\bqas
\Lambda^2 &\gg& s_{i j \dots k} = - (p_i + p_j + \dots + p_k)^2 >
(m_1 + m_2 + \dots + m_n)^2 \;\hbox{or}
\nl
\Lambda^2 &\gg& \mid t \mid \; \gg m^2\,.
\eqas
EFT mimics the unknown UV by matching the hard{-}local part of the loops, i.e.\, the terms having a bounded number
of derivatives.
Soft{-}non local components in loops cancel on both sides of the matching condition but they are not a throwaway.
The key advantage of including the non-local behavior is the appearance of some important kinematic
dependence~\cite{Donoghue:2017pgk}. Important predictions of the EFT are often related to non{-}analytic contributions
which modify tails of distributions.

\subsubsection*{Proliferation of scalars and mixing} 

The lack of discovery of beyond-the-SM (BSM) physics suggests that the SM is ``isolated'', including small mixing
between light and heavy scalars~\cite{Wells:2017aoy}: if there are many scalars then we have to assume that there is
at least the same small mixing for every one of them.

SMEFT assumes a Higgs doublet, so any mixing among scalars (in general among heavy and light degrees of
freedom) in the high{-}energy theory brings us to the HEFT/SMEFT dichotomy~\cite{Brivio:2017vri}. Although there is a
wide class of BSM models that support the (linear) SMEFT description, this realization does not always provide the
appropriate framework.

\subsubsection*{Mixing and low{-}energy limits} 

The choice of the heavy scale $\Lambda$ is crucial; in any BSM model the scale $\Lambda$ should not be confused with
the mass of some heavy degrees of freedom --  it is generally a ratio of masses and powers of couplings.
The low energy behavior of underlying theory should be computed in the mass eigenbasis, not in
the weak eigenbasis; mixing angles are function of $\Lambda$ and a large number of $1/\Lambda^2$ terms are due
to the expansion of mixing angles, not to the integration of heavy fields~\cite{Boggia:2016asg}.

\textbf{In case deviations are observed}, one needs to compare at the observable level ($\mrO$) when interpreting
the parameters of the underlying theory ($X$)~\footnote{Work in progress with A. David.}. Comparison at the $\mrO$
level implies: in $X$, with parameters $\vec{p}$, compute an observable $\mrO^i = \mrO^i_{X}(\vec{p})$.
Perform a fit of the SMEFT coefficients ($\vec{\mra}$) to a set of observables ($\vec{\mrO}$), that may but
does not need to include $\mrO^i$.
Take the best-fit coefficients ($\hat{\vec{\mra}}$) from the fit above and compute the SMEFT-predicted
observable, $\hat{\mrO^i}_{\SMEFT} = \mrO^i_{\SMEFT}(\hat{\vec{a}})$.
Perform a fit for the $X$ parameters $\vec{p}$ from the comparison to the SMEFT-predicted observable:
$\hat{\mrO^i}_{\SMEFT} \sim \mrO^i_{X}(\vec{p})$.
\textbf{Should no deviations appear to arise from the SMEFT fit to data}, $X$ comes into play.
The result of a global SMEFT fit may yield $\hat{\vec{a}}\sim 0$.
This can \textit{only} be interpreted  as ``no deviations from SM under the SMEFT assumptions'', viz.\ that of a one
doublet scalar sector.

\subsubsection*{ Linear vs. quadratic EFT representation}

Given the EFT amplitude $\mrA = \mrA^{(4)} + \frac{1}{\Lambda^2}\,\mrA^{(6)} + \frac{1}{\Lambda^4}\,\mrA^{(8)}$
``linear'' means including the interference between $\mrA^{(4)}$ and $\mrA^{(6)}$,
``quadratic'' currently means including the square of $\mrA^{(6)}$ and {\textbf{Not}}
the complete inclusion of all terms giving $1/\Lambda^4$ (even before considering $\mrA^{(8)}$). Without $\mrdim = 8$
the $1/\Lambda^4$ terms are basis{-}dependent.

In developing the (SM)EFT there is a dual motivation for including $\mrdim = 8$ operators:
it may prove that an analysis purely at the $\mrdim = 6$ level is inadequate, or that $\mrdim = 8$ effects
should be accounted for as a systematic uncertainty. Indeed, given the plethora of new physics scenarios that may
show up in future precision Higgs measurements only at the percent level, it is an urgent issue to understand and
reduce theory uncertainties so that they do not become a limiting factor in our searches for new physics.

\vspace{0.5cm}
A continuum EFT is not a model, but a sequence of low-energy effective actions
$\mrS_{\eff}(\Lambda)$, for all $\Lambda < \infty$.
EFT {\textbf{theories}}~\footnote{A theory is aimed at a generalized statement aimed at explaining a phenomenon.
A model, on the other hand, is a purposeful representation of reality.}
are being widely used in an effort to interpret experimental measurements of SM processes~\cite{Mariotti:2016owy}. In this
scenario, various consistency issues arise; one should critically examine the issues and we argue for the necessity to learn
more general lessons about new physics within the EFT approach;
inconsistent results usually attributed to the EFTs are in fact the consequence of unnecessary further approximations.


\newpage



\section*{Acknowledgements}

\noindent
JA acknowledges financial support from a grant of the Swiss National Science Foundation (Project No. P400P2\_183838). \\[1mm]
\noindent
JCC acknowledges financial support from the Spanish MECD grant FPU14,
the Spanish MINECO project FPA2016-78220-C3-1-P (Fondos FEDER) and the
Junta de Andaluc\'ia grant FQM101.\\[1mm]
\noindent
MM has been partially supported by the National Science Center, Poland,
under the research project 2017/25/B/ST2/00191, and through the HARMONIA
project under contract UMO-2015/18/M/ST2/00518.\\[1mm]
\noindent
ES has received funding  from the European Research Council
(ERC) under the European Union's Horizon 2020 research and innovation
programme (grant agreement No 788223, PanScales) and from the ERC Starting Grant PDF4BSM.\\[1mm]
\noindent
PS gratefully acknowledges financial support by the DOE (Grant No. DE-SC0009919).\\[1mm]
\noindent
DJS is supported under the ERC grant ERC-STG2015-677323.\\[1mm]
\noindent
DvD is supported by the Deutsche Forschungsgemeinschaft (DFG) within the Emmy Noether Programme under grant DY130/1-1, and through the DFG Collaborative Research Center 110 “Symmetries and the Emergence of Structure in QCD”.
The continued success of the \EOS software is contingent on the numerous contributions by physicists from all corners of the particle physics communities, to whom we express our gratitude. Special thanks go to Christoph Bobeth and Frederik Beaujean, who helped building \EOS right from the start.\\[1mm]
\noindent
AV acknowledges financial support from the Spanish grants SEV-2014-0398
and FPA2017-85216-P (AEI/FEDER, UE) and SEJI/2018/033 (Generalitat Valenciana) and the Spanish Red Consolider MultiDark FPA2017‐90566‐REDC.\\[1mm]
\noindent
JV acknowledges financial support from the European Union's Horizon 2020 research and innovation programme under the Marie Sklodowska-Curie grant agreement No 700525, `NIOBE', from the Deutsche Forschungsgemeinsahft under DFG Scientific Network, Project Number \mbox{VI 902/1-1 ``DsixTools"}, and from the Spanish MINECO through the ``Ramon y Cajal" program RYC-2017-21870. \\[1mm]
\noindent
The organizers would like to thank the IPPP Durham for all the help in the organization of SMEFT-Tools 2019. The workshop was supported by the STFC grant of the IPPP. We also thank all the participants of the workshop for their contributions and for creating a fantastic atmosphere of scientific exchange and discussions.
We hope that the next edition of the workshop will see many of our SMEFT-Tools prospects realized.

%
%
%

\renewcommand{\refname}{R\lowercase{eferences}}

\end{document}